\documentclass[trackchanges]{aastex701}

\usepackage{subcaption}
\usepackage{graphicx}
\usepackage{amsmath}
\usepackage[table,xcdraw]{xcolor}
\usepackage{multirow}

\begin{document}

\newcommand{\TP}[1]{\textcolor{blue}{#1}}

\title{The DIRECD coronal mass ejection direction catalog: three-dimensional propagation inferred from coronal dimmings}

\author[orcid=0000-0003-3883-8960]{Shantanu Jain}
\affiliation{Center for Engineering Systems and Sciences, Bolshoy Boulevard 30, bld. 1, Moscow 121205, Russia}
\email[show]{shantanu.jain@ingsci.tech}  

\author[orcid=0000-0002-9189-1579]{Tatiana Podladchikova} 
\affiliation{Center for Engineering Systems and Sciences, Bolshoy Boulevard 30, bld. 1, Moscow 121205, Russia}
\email{tatiana.podladchikova@ingsci.tech}

\author[orcid=0000-0001-5661-9759]{Karin Dissauer}
\affiliation{NorthWest Research Associates, 3380 Mitchell Lane, Boulder, CO 80301, USA}
\email{dissauer@nwra.com}

\author[orcid=0000-0003-2073-002X]{Astrid M. Veronig}
\affiliation{University of Graz, Institute of Physics, Universitätsplatz 5, 8010 Graz, Austria}
\affiliation{University of Graz, Kanzelh\"ohe Observatory for Solar and Environmental Research, Kanzelh\"ohe 19, 9521 Treffen, Austria}
\email{astrid.veronig@uni-graz.at}

\author[0000-0003-2735-5832]{Amaia Razquin}
\affiliation{University of Graz, Institute of Physics, Universitätsplatz 5, 8010 Graz, Austria}
\email{amaia.razquin-lizarraga@uni-graz.at}

\begin{abstract}
	
Coronal mass ejections (CMEs) are among the primary drivers of space weather disturbances at Earth, yet their early propagation in the low corona remains poorly constrained owing to occultation and projection effects inherent to coronagraph observations. Coronal dimmings offer an alternative diagnostic to infer CME propagation direction directly from the low corona. We present the DIRECD CME Direction Catalog, a unified dataset of three-dimensional CME propagation directions derived via the DIRECD (Dimming Inferred Estimation of CME Direction) method, which reconstructs CME cone geometry from SDO/AIA 211 \AA\ coronal dimming observations. The catalog comprises 64 events spanning 2010-2026, combining 31 Solar Cycle 25 (2021-2026) events with 33 events from prior studies. For each event we report the 3D propagation direction, 2D inclination angles, angular width, and cone height. Statistical analysis reveals a systematic asymmetry in low-coronal CME propagation: meridional inclinations exhibit a systematic tendency toward latitude-dependent deflection, with poleward deflections more pronounced at higher source latitudes, while equatorial inclinations show no significant dependence on source longitude. This is consistent with low-coronal CME trajectories being governed primarily by local active-region magnetic topology, with the large-scale coronal field and heliospheric current sheet assuming a progressively dominant role at greater heliocentric distances. Incorporating DIRECD-derived propagation directions into CME arrival forecasts improves predictions of geomagnetic storm intensity, raising the correlation with observed storm strength from $r = 0.58$ to $r = 0.70$ when combined with CME speed. The DIRECD software, catalog, and all associated data products are publicly available to support space weather research.

\end{abstract}


\section{Introduction}\label{intro}
Coronal mass ejections (CMEs) are large-scale eruptions of magnetized plasma from the Sun, 
propagating at speeds up to several thousand kilometers per second and are major 
drivers of space weather \citep{michalek2009expansion, gopalswamy2009soho, tsurutani2014extreme, Cheng2017, Veronig2018_Genesis}. Predicting CME arrival times and impacts depends critically on accurate knowledge of the CME propagation direction. In particular, CME impact at Earth, arrival time and speed predictions depend on the assumed propagation 
direction in heliospheric propagation and ensemble forecasting models. Multiple operational and research-oriented forecasting approaches require early CME direction estimates as input, including cone-based ensemble models, drag-based ensemble methods (DBEM) \citep{vrvsnak2013propagation, dumbovic2018drag, abu2025coronal}, and physics-based magnetohydrodynamic (MHD) models such as ENLIL \citep{luhmann2017modeling}, EUHFORIA \citep{pomoell2018euhforia}, SUSANOO-CME \citep{shiota2016magnetohydrodynamic}, which require CME direction as a direct input parameter, as well as global MHD frameworks such as AWSoM \citep{van2014alfven} within the Space Weather Modeling Framework, where direction information instead constrains the location and orientation of the inserted flux. For example, ELEvoHI requires the propagation direction and angular 
width as key input parameters \citep{Rollett2012,Rolett2016,Amerstorfer2021}.  Beyond Earth, propagation direction determines whether other planetary environments are impacted, an important consideration when interpreting CME and shock related particle and radiation signatures at Mars and along interplanetary trajectories \citep{Guo2018, FreiherrvonForstner2019}. Finally, CME direction and the resulting magnetic connectivity can influence the access of observers to CME-driven shocks and the observed longitudinal spread and timing of SEP events. As such, constraining CME direction is equally important in SEP modeling and event analysis contexts \citep{Palmerio2021,Dresing2023}.

Several three-dimensional reconstruction techniques have been developed to investigate the propagation, geometry, and trajectory of coronal mass ejections (CMEs), including their deflection within the heliosphere. Among the most widely used is the Graduated Cylindrical Shell (GCS) model \citep{Thernisien2006, Thernisien2011}, a forward-fitting method that reconstructs the large-scale structure of CMEs from stereoscopic coronagraph images, typically acquired at heliocentric distances beyond a few solar radii. Additional multi-viewpoint approaches include elliptical tie-pointing \citep{Byrne2010} and geometric triangulation \citep{Liu2010, Podladchikova2019three}, both of which constrain CME kinematics and direction by leveraging simultaneous observations from widely separated spacecraft. These observations rely on coronagraph observations in the outer corona.  

However, coronagraphs are limited by the 
occulting disk, which obstructs the view of the low corona where CMEs originate \citep{burkepile2004role, schwenn2005association}. This makes 
it difficult to determine the early direction of CME propagation. However, alternative observational 
signatures in the low corona, such as coronal dimmings, can provide early direction constraints.
Coronal dimmings are localized reductions in extreme ultraviolet (EUV) and soft X-ray 
emissions in the low corona, caused by plasma depletion during CME eruptions \citep{hudson1996long, sterling1997yohkoh, thompson1998soho}. For a comprehensive discussion on dimming properties and their links to CMEs,  we refer to the recent review by \cite{Veronig2025LRSP}.

Extensive research has established a close association between dimming signatures and CME properties, including mass, speed and morphological evolution \citep{harrison2000spectroscopic, harrison2003coronal, zhukov2004nature, lopez2017mass, Dissauer2018a, Dissauer2018b, Dissauer2019, chikunova2020coronal, attrill2006using, qiu2017gradual, razquin2025coronal}. Coronal dimmings may also serve as a proxy for stellar CMEs \citep{Jin2020coronal, Veronig2021indications}. 

A new approach, the DIRECD (Dimming 
Inferred Estimation of CME Direction) method, exploits coronal dimming geometry to infer 
three-dimensional CME propagation directions from observations in the low corona, before 
CME structures become fully visible in coronagraph data. The DIRECD method has been 
validated against the GCS model reconstructions and coronagraph observations, demonstrating 
that dimming observations reliably constrain early CME direction \citep{jain2025validating, jain2024coronal, jain2024estimating, podladchikova2024three}.

In this paper, we present and discuss in detail 31 newly analyzed CME events that occurred during solar cycle 25, while publicly releasing the DIRECD CME direction catalog, which additionally includes 33 CME events adopted from earlier DIRECD statistical studies \citep{jain2025validating}, resulting in a total of 64 cataloged CME/dimming events. The catalog provides a complete and standardized set of coronal dimming masks and timing information for all events. The detailed analysis and validation presented in this paper focus on the newly analyzed Solar Cycle~25 events, for which we derive and discuss the full set of DIRECD-derived CME geometry parameters, including the 3D propagation direction, inclination angle, angular width, and the cone height at which the erupting structure remains connected to the dimming footprint, together with associated uncertainty estimates derived in this study. The legacy events are included in the catalog as standardized dimming data products, enabling direct reuse and comparison, while their corresponding DIRECD results are documented in earlier analyses. Minor numerical differences may arise as the software continues to evolve and detection routines are refined.

All data products are released as a unified dataset across solar cycles 24-25 for the time range 2010-2026 that can be used as a direct input for DIRECD, for studies of early CME properties. The catalog is generated using the newly developed open-source DIRECD software available on Github\footnote{\url{https://github.com/jain-shantanu/DIRECD}} and Zenodo\footnote{\url{https://doi.org/10.5281/zenodo.20487897}}, the primary objective of which is to provide the solar and heliospheric physics community with a robust, user-friendly platform for automated dimming detection and early CME direction estimation using the DIRECD method.  We demonstrate for the first time that DIRECD operates in 
near real-time on SDO quick-look data, showing applicability for operational space 
weather forecasting. The SDO quick-look data consists of near-real-time AIA Level 1 data (JSOC series aia.lev1\_nrt2).

This paper is organized as follows. Section~\ref{case_study} presents two representative CME events - 24 February 2023 and 18 January 2026 and demonstrates the full DIRECD analysis workflow, from automated dimming detection to three-dimensional CME direction reconstruction and validation with coronagraph observations. Section~\ref{CME_catalog} describes the construction, content, and statistical properties of the DIRECD CME direction catalog, including uncertainty estimation, cross-validation for an event, and the released data products. Section~\ref{conclusions} summarizes the key findings and outlines future directions for extending the DIRECD method toward operational and off-limb applications.


\section{Representative Catalog CME Events: DIRECD Analysis of 24 February 2023 and 18 January 2026 events}\label{case_study}
The DIRECD method uses a 3D cone model to approximate an expanding CME at the end of the dimming impulsive phase and to infer key CME parameters, including the propagation direction, half-width, and the cone height at which the CME remains connected to the dimming, by matching the cone projections on the solar sphere to the observed dimming geometry \citep{jain2024estimating}. To illustrate the method, we present a partial-halo CME (24 February 2023) event, and recent extreme CME (18 January 2026) halo event which are included in the DIRECD CME direction catalog.

The 24 February 2023 CME was associated with an M3.7-class solar flare, as recorded by the GOES X-ray flux measurements and cataloged in the GOES/HINODE flare catalog \citep{watanabe2012hinode}. The accompanying coronal mass ejection (CME) was identified in the LASCO CME catalog \citep{2004JGRA..109.7105Y}, with a speed of $\sim 1336$ km~s$^{-1}$ in the LASCO coronagraph field of view. The CME struck earth on 26 February 2023, triggering a G1-G2 magnetic storm\footnote{\url{https://www.spaceweather.com/archive.php?view=1&day=26&month=02&year=2023}}. This interaction produced a moderate geomagnetic storm, reaching a peak Dst index of –132~nT on 27 February 2023\footnote{\url{https://omniweb.gsfc.nasa.gov}}. In contrast, the 18 January 2026 CME was accompanied by an X1.9-class solar flare and was a very fast halo CME, propagating at approximately speed of 1842~km~s$^{-1}$ as identified in the LASCO CME catalog \citep{2004JGRA..109.7105Y}. The CME traversed the Sun–Earth distance in about 25 hours and impacted Earth nearly head-on \footnote{\url{https://www.spaceweather.com/archive.php?view=1&day=19&month=01&year=2026}}, triggering a G4-class geomagnetic storm that reached a minimum Dst index of –214 nT on 20 January 2026 at 17:00~UT inspected with OMNI database \citep{papitashvili2020omni}, successfully predicted several hours in advance by the StormFocus service \citep{Podladchikova2012,Podladchikova2018}\footnote{\url{https://spaceweather.ru/content/extended-geomagnetic-storm-forecast}}, as shown in the prediction screenshot\footnote{\url{http://www.iki.rssi.ru/forecast/data/Archive/2026/en/01/20/Prediction_20260120_2305.gif}}, and associated with the most intense solar proton event recorded in the 21st century to date.

To generate all figures and data products associated with this event, including the coronal dimming masks and estimated 3D CME geometry we use the newly developed open-source DIRECD software\footnote{\url{https://zenodo.org/records/20487897}}, released as part of the DIRECD catalog. The software provides automated routines for coronal dimming detection, CME cone reconstruction, uncertainty estimation, and visualization of results. Detailed information on the software architecture, installation, input/output formats, and user-level operation is provided in the accompanying online documentation and user manual\footnote{\url{https://direcd.readthedocs.io/en/latest/}}.

We identify coronal dimmings associated with the CME events using automated detection methods \citep{Dissauer2018a,Chikunova2023} applied to SDO/AIA (Solar Dynamics Observatory/Atmospheric Imaging Assembly) 211~\AA~observations \citep{lemen2012atmospheric,pesnell2012solar}. All SDO/AIA images are processed using the SunPy library \citep{mumford2015sunpy}, which involved selecting exposure times ($>1$~s), correcting for differential rotation, and resampling to a uniform $1024 \times 1024$ pixel resolution. The procedure constructs logarithmic base-ratio images relative to a pre-event reference time (30 min before the flare) and segments dimming regions through adaptive thresholding combined with a region-growing algorithm to derive instantaneous and cumulative dimming masks.  A variable threshold between $–0.11$ and $–0.19~DN$ is used to segment dimming regions based on histogram thresholding, which are subsequently processed with morphological operations and median filtering ($3 \times 3$ pixel kernel) to suppress noise. For the two example events, we use the mean value of $-0.15~DN$ for segmentation, which was selected by visual analysis of dimming development. The detection of coronal dimmings is performed within a $1000" \times 1000"$ subfield centered on the eruption site. The detected dimming pixels are accumulated in time to construct a cumulative dimming map, and the dimming area is calculated by estimating the surface area of the solar sphere for every pixel following the approach of \citet{Chikunova2023}, which accounts for projection effects, particularly near the solar limb.

\begin{figure}[h]
	\centering 
	\subfloat[]{%
		\includegraphics[width=0.95\columnwidth]{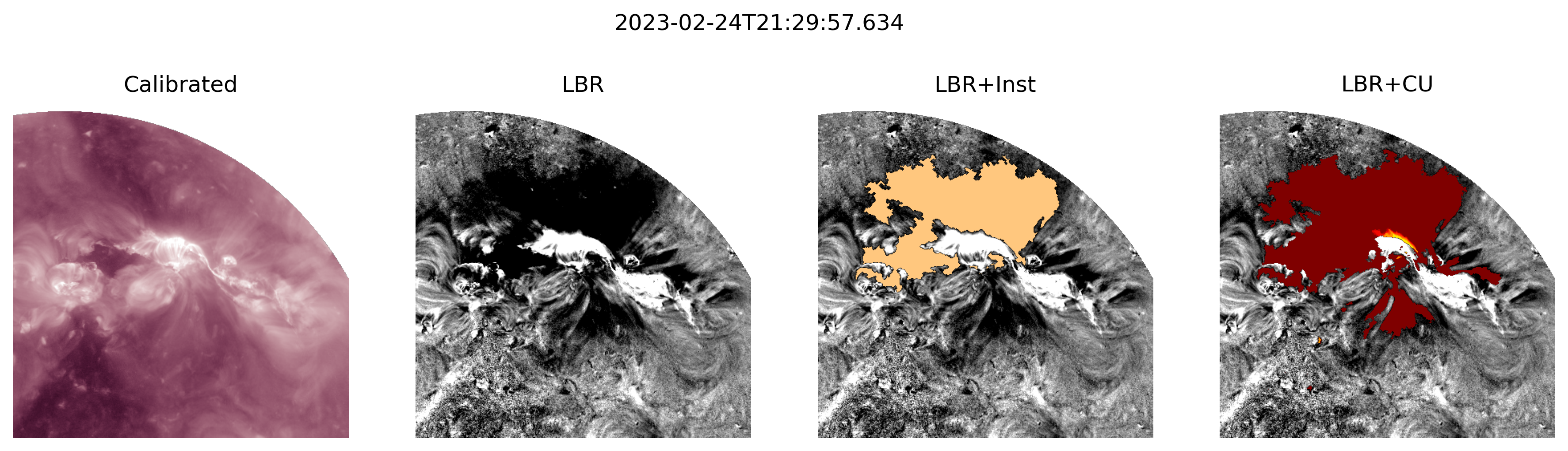}%
	}\vspace{0.2cm}
	\subfloat[]{%
		\includegraphics[width=0.95\columnwidth]{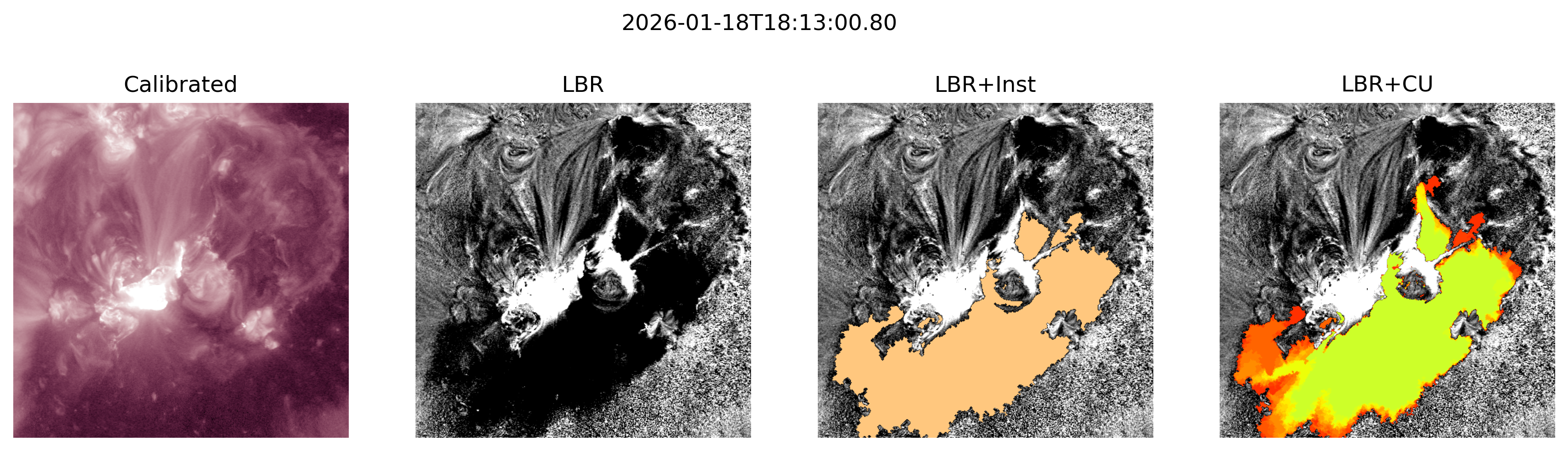}%
		
	}
	\vspace{-0.2cm}
	\includegraphics[width=0.9\columnwidth]{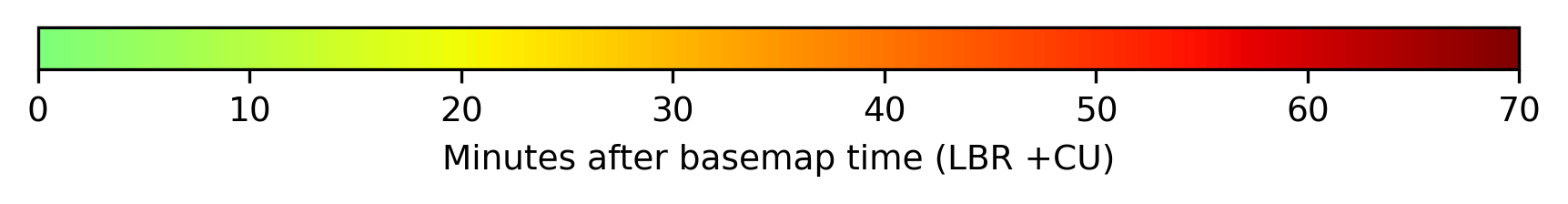}
	\vspace{-0.2cm}
\caption{Dimming detection for the 24 February 2023 event (a) and the 18 January 2026 event (b) with the DIRECD software. The 4 panels in a row show the calibrated SDO/AIA image, the logarithmic base-ratio image, the instantaneous and the cummulative dimming detection on the logarithmic base-ratio image respectively. In the LBR+CU panels, the colormap is based on the time at which each pixel was first detected as dimming relative to the base image}.
\label{detection_soft}
\end{figure}

Following the approach established in previous studies \citep{Dissauer2018b, jain2024coronal, jain2024estimating, jain2025validating}, we define the end of the dimming impulsive phase as the time when the cumulative dimming area growth rate decreases to 15\% of its maximum value. The cumulative dimming mask at the end of this phase serves as the primary observational constraint for reconstructing the CME geometry. To identify the dominant direction of dimming expansion, we divide the solar disk into angular sectors centered on the source region and evaluate the cumulative dimming area in each sector \citep{jain2024coronal}. Figure~\ref{detection_soft} shows the dimming detection map at the end of the impulsive phase for the two events under study. For each event we show 4 plots in a row: the calibrated SDO/AIA image, the logarithmic base-ratio (LBR) image, the instantaneous dimming detection superimposed on LBR image and the cumulative dimming detection map on LBR image respectively.

\begin{figure}[h]
	\centering 
	\subfloat[]{%
		\includegraphics[width=0.46\columnwidth]{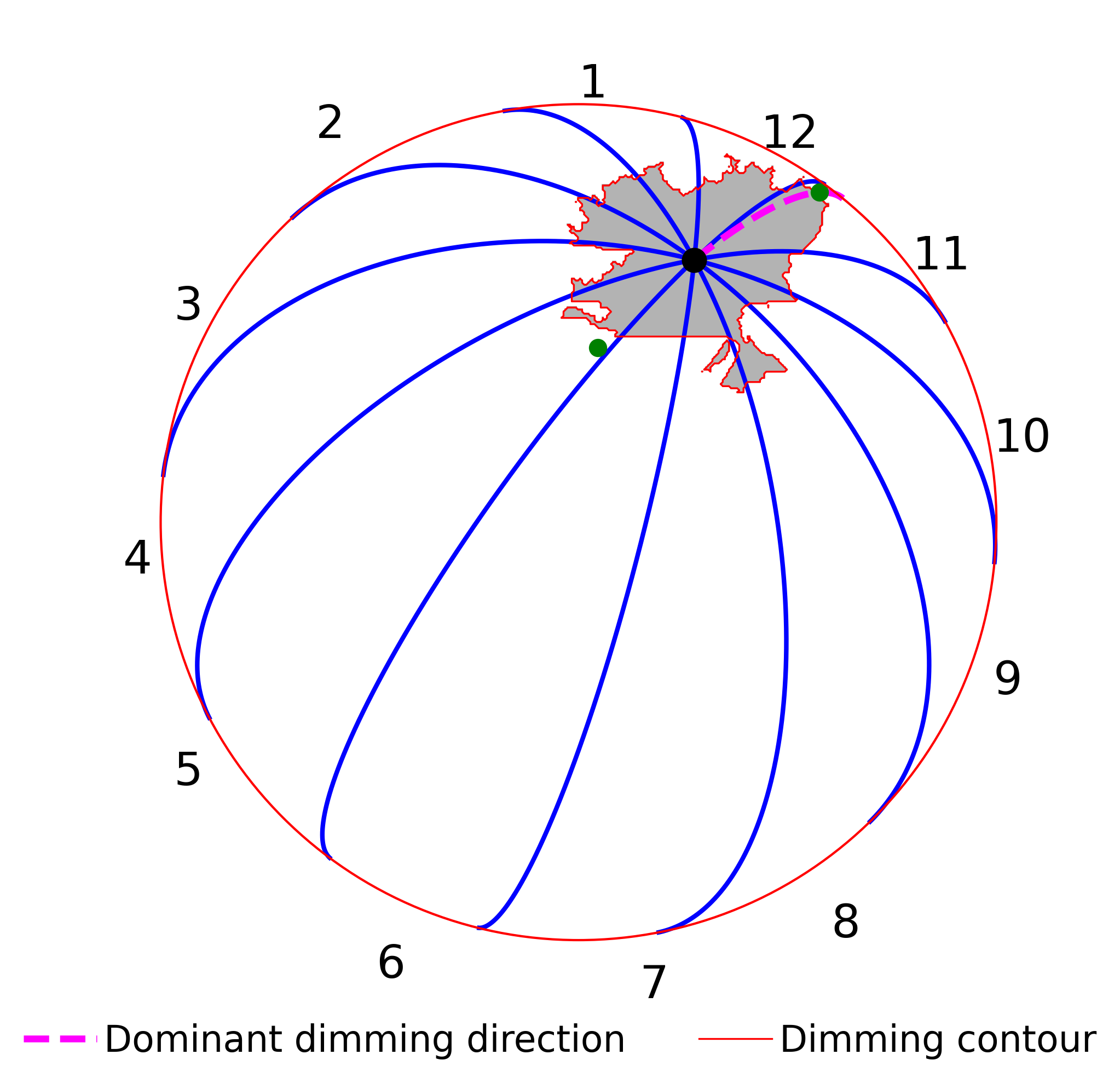}%
		\label{dimming_edges_1}
	}\qquad
	\subfloat[]{%
		\includegraphics[width=0.46\columnwidth]{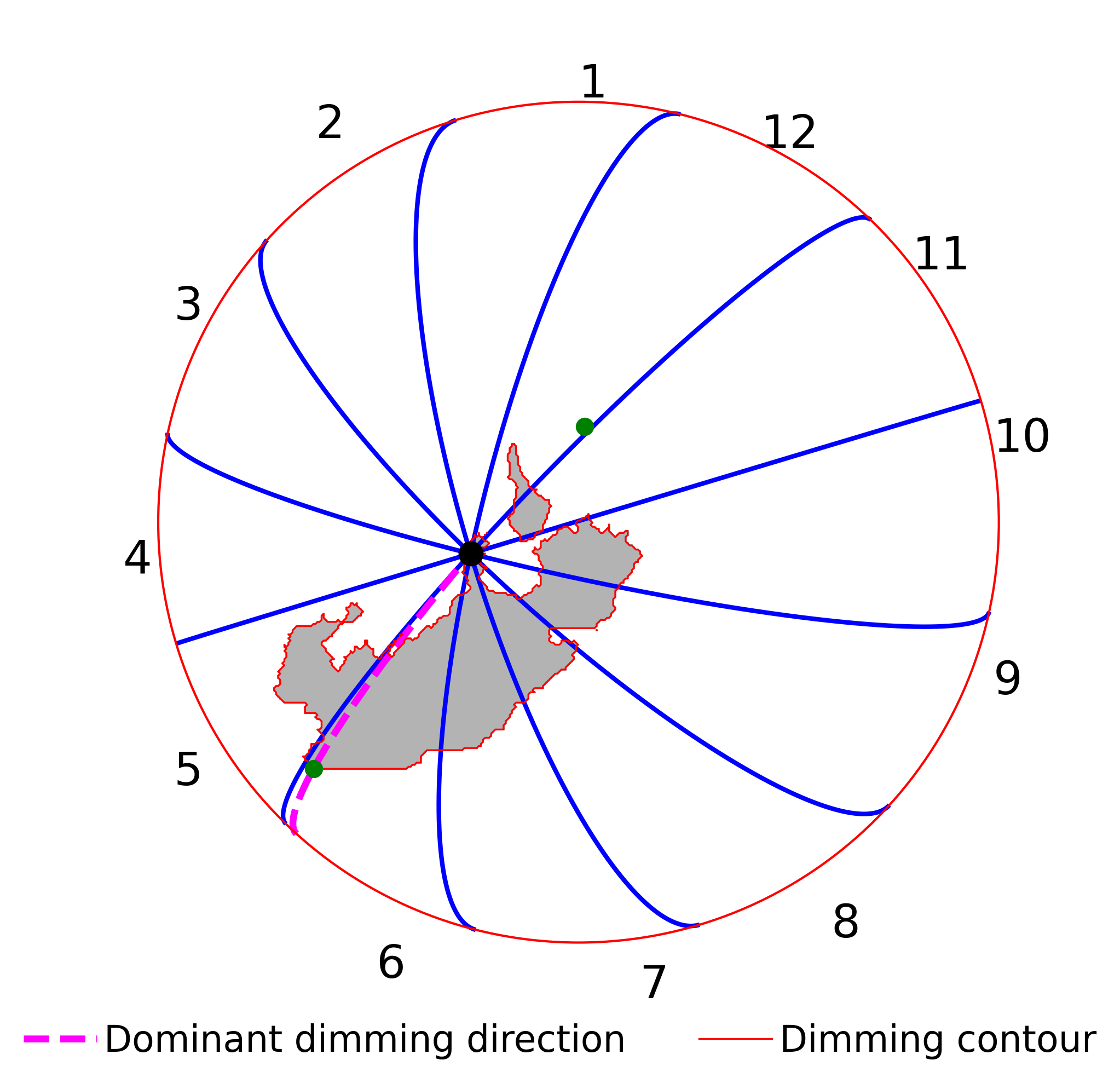}%
		\label{dimming_edges_2}
	}
	\caption{Determination of dimming edges and dominant dimming direction from the cumulative dimming mask plotted at the end of the impulsive phase. The dominant dimming direction is in sector 11 for (a) 24 February 2023 event and in sector 6 for (b) 18 January 2026 event. The green dots show the edges of dimming for cone generation and the pink curve shows the largest dimming extent in the sector of dominant dimming development}
		
	\label{dimming_edges}
\end{figure}

We note that the 18 January 2026 event exhibits an unusually short impulsive phase (13 minutes) compared to the other events (Table \ref{table:events} in section \ref{Event selection and data processing}), consistent with it being the fastest event in our sample (X1.9 flare, ~1842 km/s CME speed). At the 1-minute cadence used throughout this study, the impulsive phase for this event is sampled by ~13 data points, sufficient to identify the 15\% area-growth-rate threshold but coarser in time resolution than for the longer-duration events in the catalog. We verified that the cumulative dimming mask at the identified end-of-impulsive-phase time is not sensitive to the precise choice of threshold within the adjacent one-to-two cadence steps for this event, supporting the robustness of the derived CME geometry despite the short phase duration.
Figure \ref{dimming_edges} shows the angular sector scheme and cumulative dimming contours 
for both events. The blue radial lines indicate the 12 equally-spaced sectors, and the 
gray shaded regions show the cumulative dimming extent at the end of the impulsive phase. For the 24 February 2023 event (Figure \ref{dimming_edges_1}), the cumulative dimming area reaches its maximum in sector 11, indicating the dominant dimming, and thus CME propagation direction. For the 18 January 2026 event, the dominant dimming direction is in sector 6 (Figure \ref{dimming_edges_2}).

Using this dominant direction and the dimming edges that define the surface extent of the CME footprint, we construct an ensemble of CME cones with varying heights, angular widths, and inclinations relative to the radial direction. For each cone, we compute orthogonal projections onto the solar sphere and compare the resulting projected footprints with the observed dimming extent at the end of the impulsive phase. We select the CME parameters by identifying the cone configuration that best reproduces the dimming geometry while maximizing the fraction of dimming pixels enclosed within the projected cone boundaries \citep{jain2025validating}. This procedure results in an estimate of the CME propagation direction in 3D, together with associated geometric parameters such as inclination angle,  angular width, and CME cone height, at which the CME remains connected to the dimming. 

\begin{figure}[h]
	\centering 
	\subfloat[]{%
		\includegraphics[width=0.45\columnwidth]{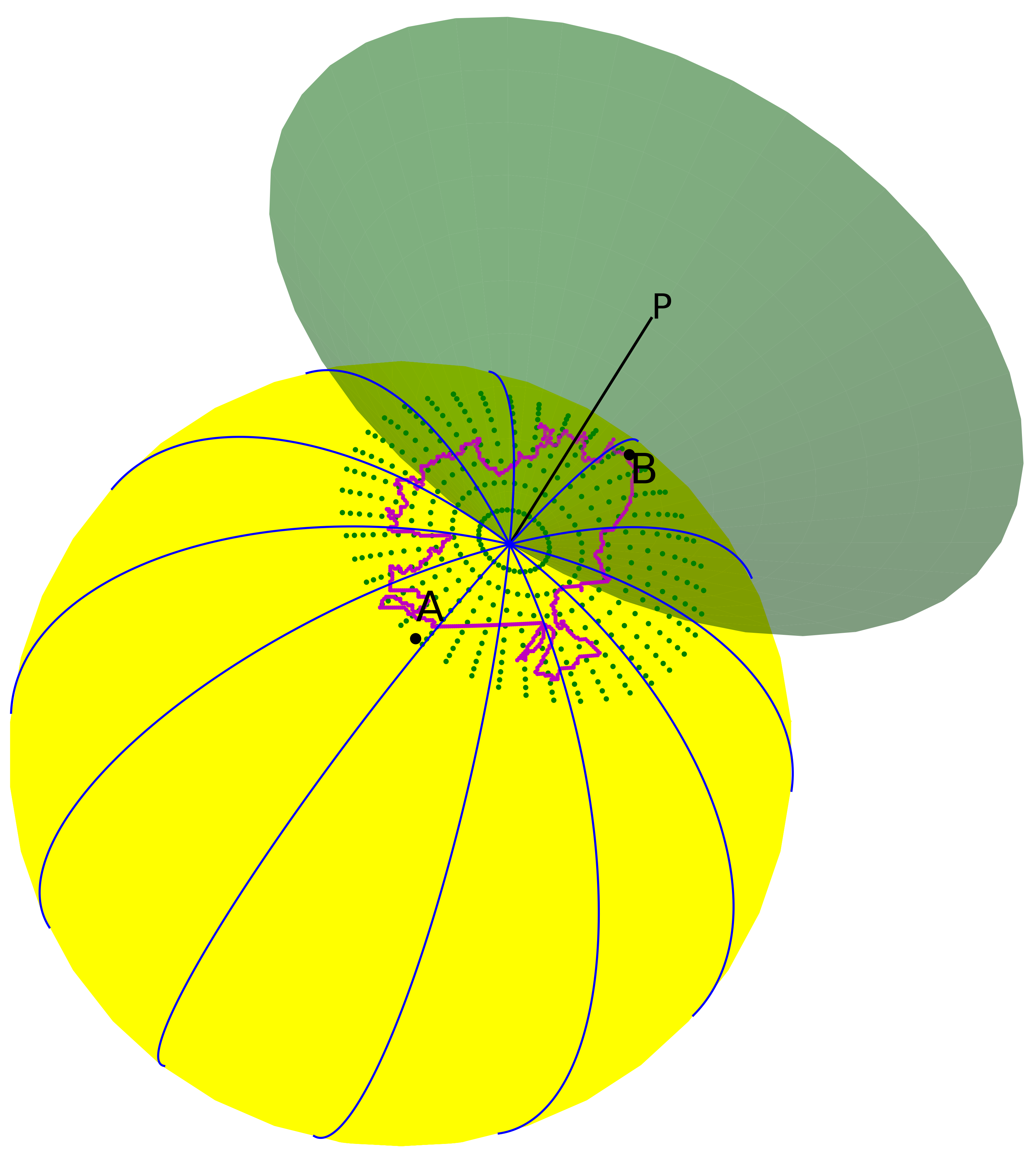}%
		\label{24022023_3d_cone}
	}\qquad
	\subfloat[]{%
		\includegraphics[width=0.45\columnwidth]{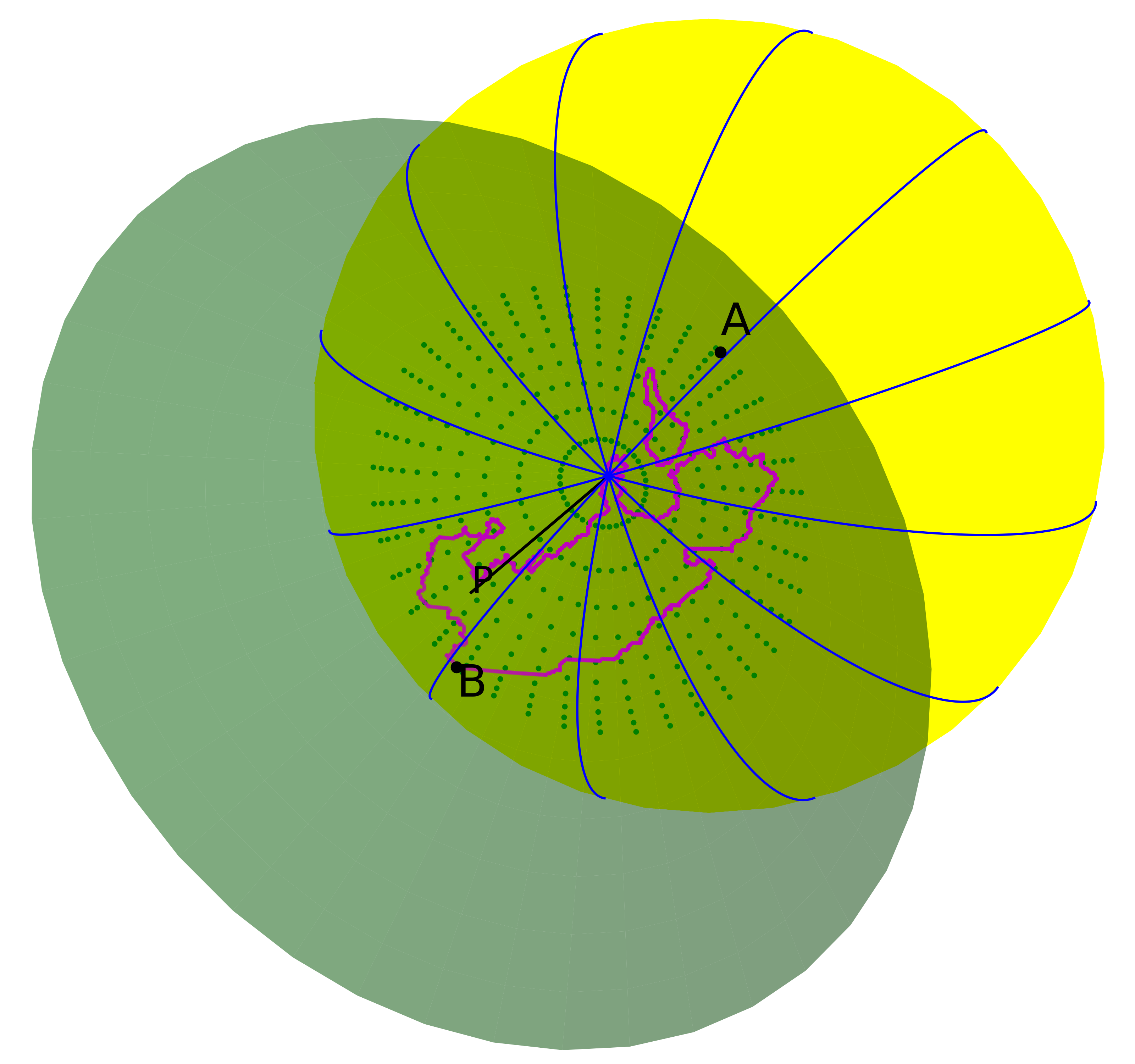}%
		\label{18012026_3d_cone}
	}
	\caption{Best-fit DIRECD cone reconstruction overlaid dimming for (a) 24 February 2023 event and (b) 18 January 2026 event. The 
		three-dimensional cone model represents the best-fit CME geometry that reproduces the observed dimming extent (magenta boundaries). The cone's angular width, height, and inclination are derived to match the observed dimming pattern. Points A and B mark the locations of maximum dimming extent, P marks the apex of the cone's main axis, and green dots represent the orthogonal 	projections of the cone footprint onto the solar disk. Derived parameters are 
		listed for each event: (a) cone height =  $0.95 R_{\sun}$, inclination angle = $15.9^\circ$, 
		angular width = $99.7^\circ$; (b) cone height = $1.0 R_{\sun}$, inclination angle = \(13.2^\circ\), 
		angular width = $93.2^\circ$.
		}
	\label{3D_cones}
\end{figure}

The resulting best-fit cones for both events are shown in Figure~\ref{3D_cones}. For the 24 February 2023 event, the reconstruction yields a cone height of $0.95,R_{\sun}$, an angular width of $99.6^\circ$, an inclination angle of $15.9^\circ$ relative to the solar radial direction, with the cone axis tilted $11.9^\circ$ toward the north and $14.6^\circ$ toward the west. For the 18 January 2026 event, the best-fit cone has a height of $1.0,R_{\sun}$, an angular width of $93.2^\circ$, an inclination angle of $13.2^\circ$, with the axis inclined $8.8^\circ$ toward the south and $12.9^\circ$ toward the east.

\begin{figure}
	\centering 
	\subfloat[]{%
		\includegraphics[width=0.45\columnwidth]{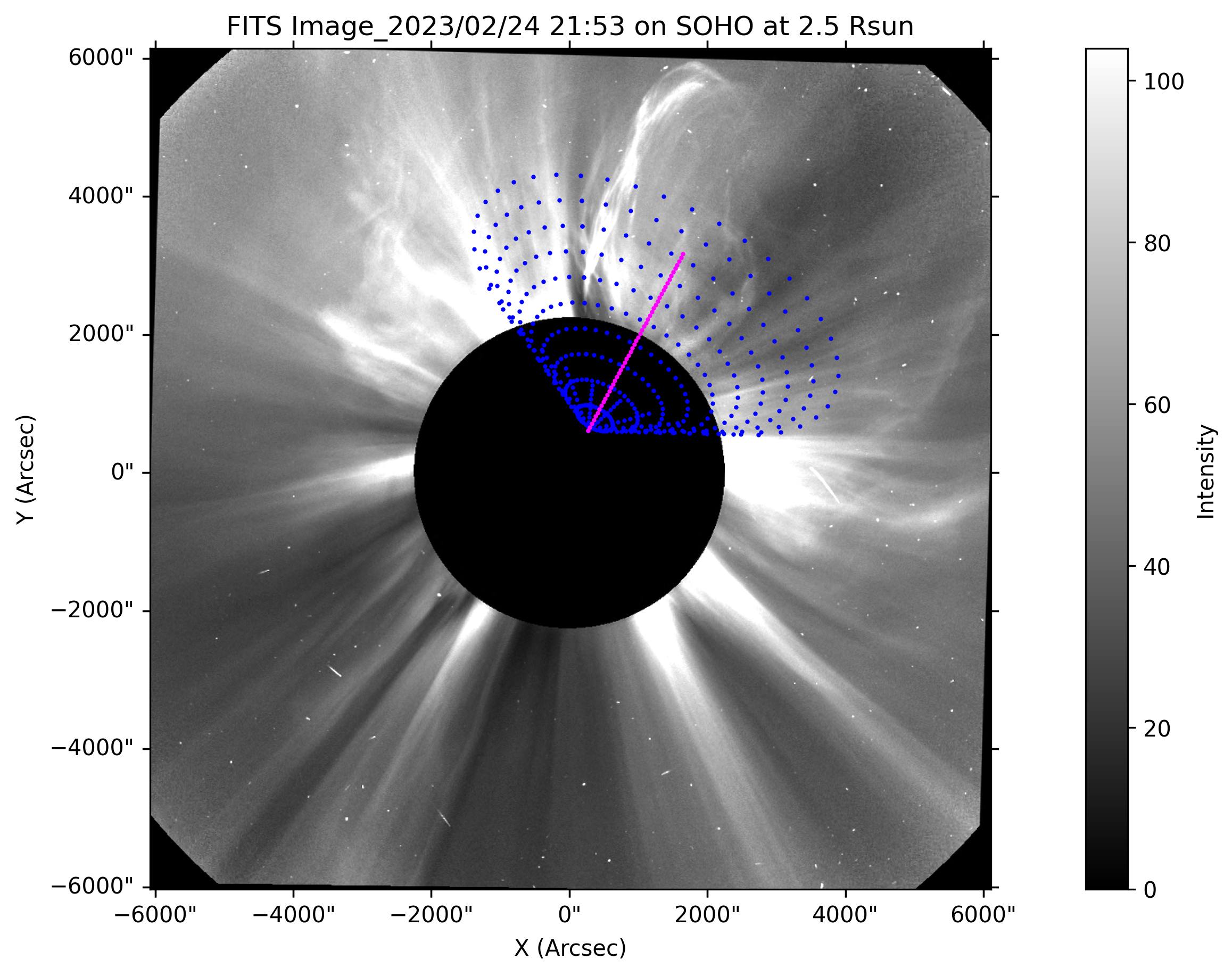}%
	}\qquad
	\subfloat[]{%
		\includegraphics[width=0.45\columnwidth]{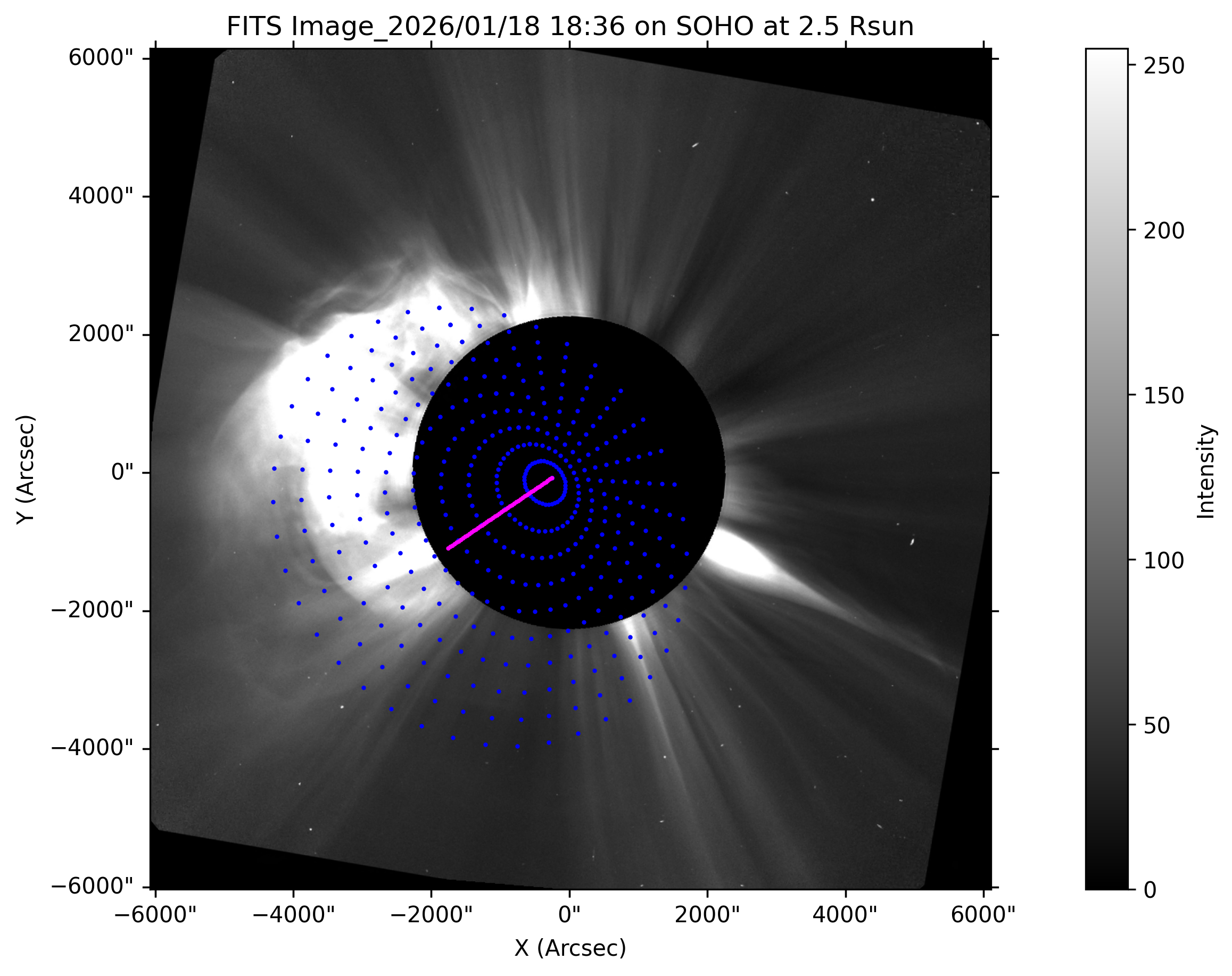}%
		
	}\\
	\subfloat[]{%
		\includegraphics[width=0.45\columnwidth]{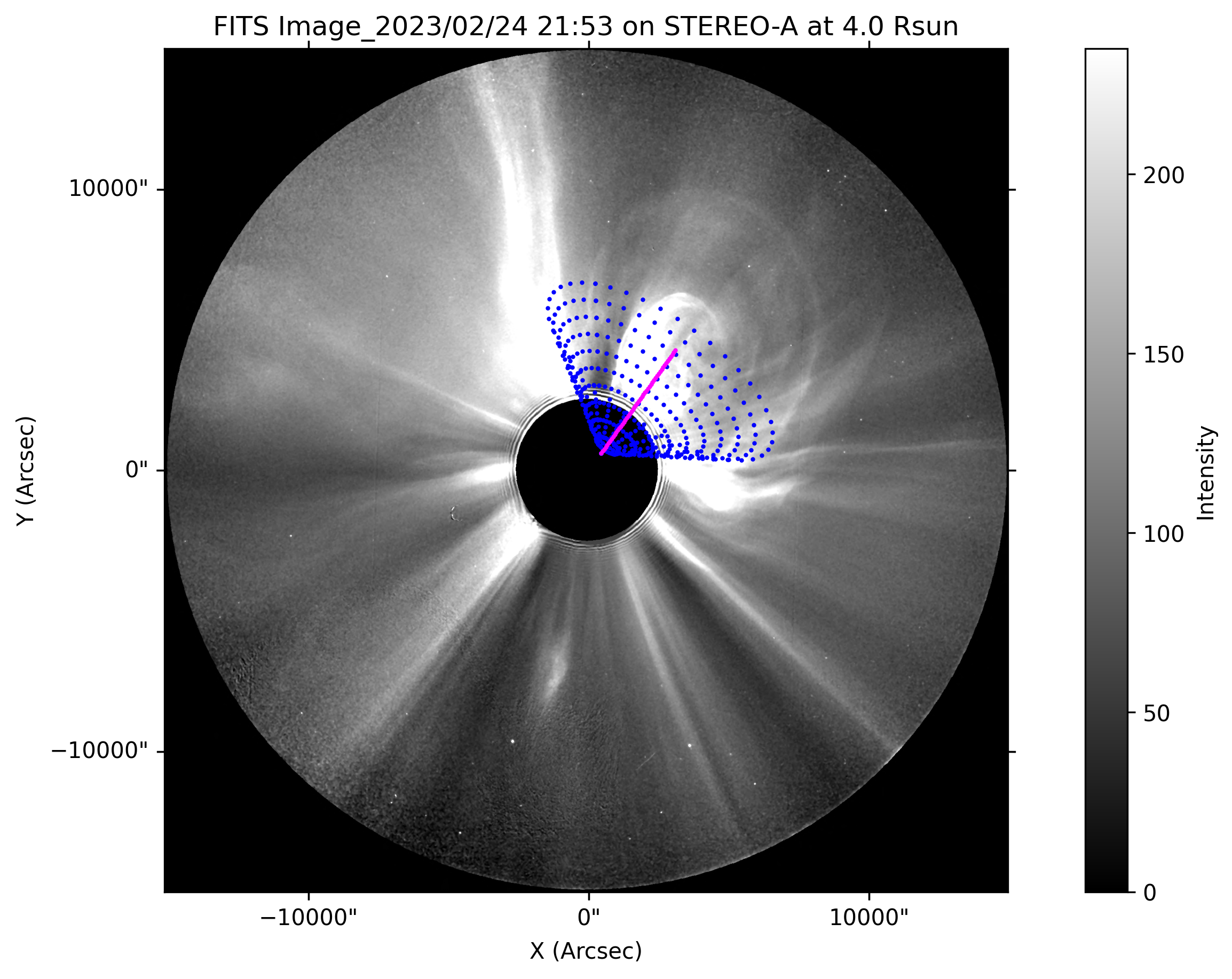}%
	}\qquad
	\subfloat[]{%
		\includegraphics[width=0.45\columnwidth]{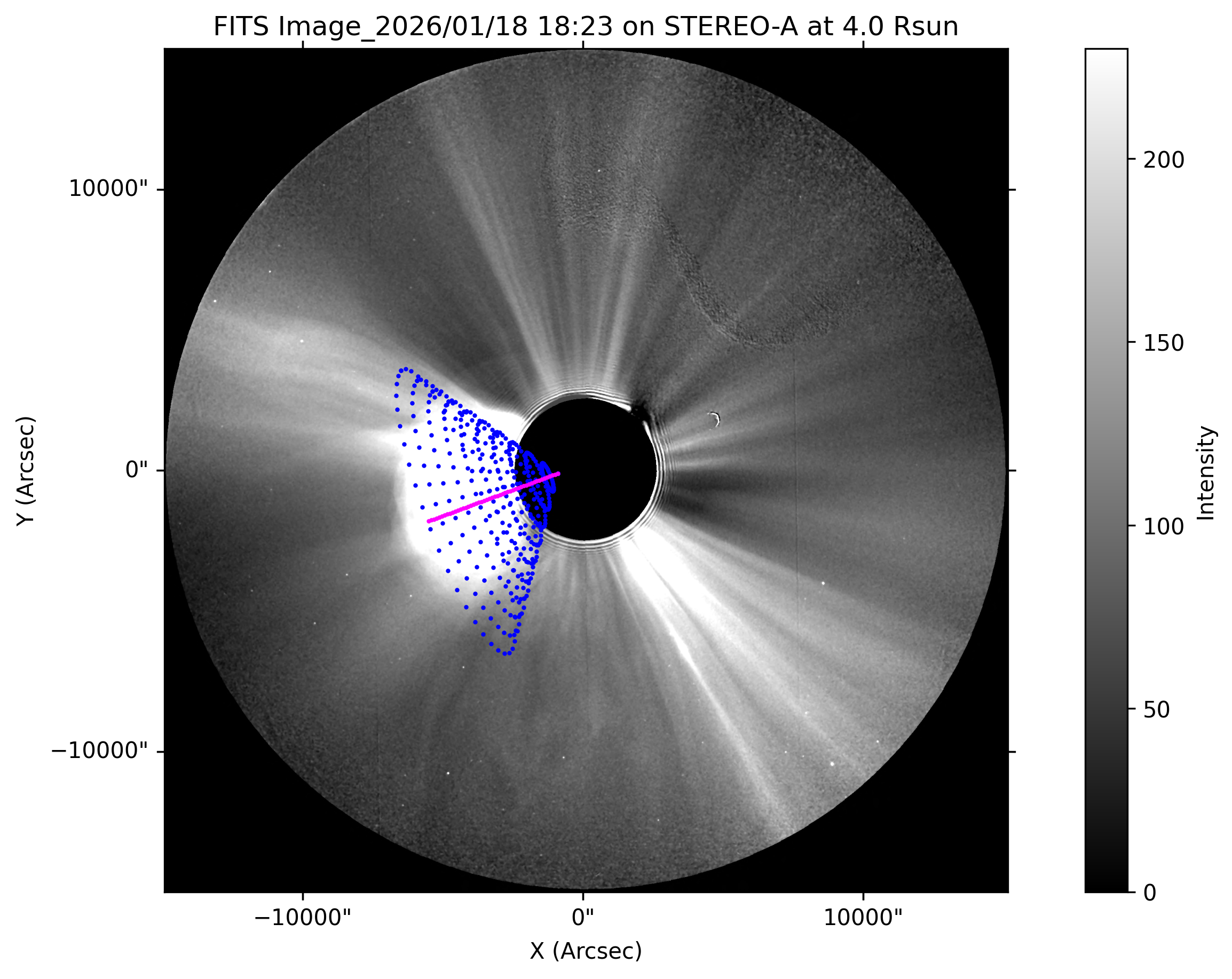}%
		
	}
	\caption{	
		Best fit CME cone projections derived from DIRECD overlaid on LASCO/STEREO-A images. The cones at extended to heights \(2.5 R_{\sun}\) and \(4 R_{\sun}\) respectively for plotting on LASCO/STEREO images. Panels (a) and (b) show the best fit cone projections extended to heights \(2.5 R_{\sun}\) on LASCO/C2 coronagraph for 24 February 2023 and 18 January 2026 events respectively. Panels (c) and (d) show the best fit cone projections extended to heights \(4 R_{\sun}\) on STEREO/COR-2A coronagraphs for 24 February 2023 and 18 January 2026 events respectively. }
\label{validation}
\end{figure}

We plot the CME directions inferred from coronal dimmings on the white-light coronagraph observations from SOHO/LASCO. The projected CME cones derived from the dimming analysis are overlaid onto LASCO and STEREO images, enabling a direct comparison between the low-coronal propagation directions inferred from dimmings and the CME morphology seen higher in the corona (Figure~\ref{validation}). For both events, the cone projections are oriented towards the center of the CME bubble in LASCO and STEREO-A observations, providing confidence that the dimming-inferred directions reliably capture the early CME propagation.

\section{The DIRECD CME direction catalog}\label{CME_catalog}
The example above illustrates the full DIRECD workflow applied to two events. In the following section, we extend this analysis to a systematically selected set of CME events, forming a homogeneous catalog of dimming-derived CME propagation directions.

\subsection{Event selection and data processing} \label{Event selection and data processing}

The catalog is based on dimming evolution derived from full-disk SDO/AIA EUV observations. We use the 211~\AA~band because it provides strong dimming contrast at relevant coronal temperatures \citep{Dissauer2018a, kraaikamp2015solar}. For each event, we analyze a 2.5-hour interval sampled at a one-minute cadence, starting 30 minutes before the associated flare onset. We preprocess the data and detect coronal dimmings using the same pipeline described in Section~\ref{case_study}; for consistency across the catalog, we apply the detection within a $1000\times1000$~arcsec subfield around the source region and use LBR thresholding in the range of $-0.11$ to $-0.19$~DN.

The catalog includes 31 CME events from Solar Cycle 25 selected according to a set of uniform criteria designed to ensure reliable dimming identification and robust CME association. The event selection criteria is similar to the one outlined in \citep{jain2025validating}. Each of the events must exhibit well-defined, coherent coronal dimming signature in SDO/AIA 211~\AA~base-ratio images. Moreover, each dimming event must be associated with a confirmed solar flare with clear onset time and source coordinates, verified using the GOES flare 
catalog \footnote{http://www.ngdc.noaa.gov/stp/satellite/goes/index.html}, the HINODE flare catalog \citep{watanabe2012hinode}, or NOAA's 
Space Weather Prediction Center records \footnote{\url{https://www.swpc.noaa.gov/products/goes-x-ray-flux}}. Additionally, the associated flare's source region is constrained to within $\pm 60^\circ$ (potentially upto  $\pm 75^\circ$ depending on dimming morphology) of the solar disk center to minimize projection effects in the dimming identification process.

Columns 1-6 of Table \ref{table:events} gives an overview of the 31 newly analyzed Solar Cycle 25 events and their associated dimming parameters such as the date of the event, start and end of the impulsive phase of the dimming and the heliographic location of the associated flare.

\begin{table}[htbp]
	\centering
	\raggedright
	\small
	\caption{List of selected dimming events and DIRECD results.}
		\begin{tabular}{|c|l|c|c|c|c|c|c|c|c|}
			\hline
			\textbf{\#} & \textbf{Event Date} & \textbf{Start Time} & \textbf{End Time} & \textbf{Lat. } & \textbf{Lon.} & \textbf{Best Height} & \textbf{Best Incl.} & \textbf{Meridional} & \textbf{Equatorial} \\
			& & (UTC) &(UTC) &  ($^\circ$) & ($^\circ$) & ($R_{\sun}$) & ($^\circ$) & ($^\circ$) & ($^\circ$) \\ \hline
			1 & 28-09-2021 & 05:23 & 07:52 & $-28$ & 40 & 0.85 & 8.1 $\pm$ 1.8 & 3.3 $\pm$ 0.7 & 8.1 $\pm$ 1.7 \\ \hline
			2 & 09-10-2021 & 06:20 & 07:07 & 19 & $-7$ & 0.93 & 12.7 $\pm$ 1.9 & 4.1 $\pm$ 0.6  & 13.1 $\pm$ 1.8 \\ \hline
			3 & 02-11-2021 & 01:20 & 02:59 & 23 & $-7$ & 0.67 & 6.3 $\pm$ 1.9 & 5.9 $\pm$ 1.8 & $-2.2$ $\pm$ 0.7 \\ \hline
			4 & 14-01-2022 & 12:17 & 13:40 & $-35$ & 35 & 0.80 & 0.2 $\pm$ 1.6 & 0.0 $\pm$ 0.1 & $-0.2$ $\pm$ 1.6 \\ \hline
			5 & 28-03-2022 & 10:46 & 11:45 & 16 & 4 & 1.10 & 2.5 $\pm$ 1.6 & 2.4 $\pm$ 1.6 & 0.1 $\pm$ 0.1 \\ \hline
			6 & 30-03-2022 & 17:21 & 18:38 & 13 & 32 & 1.15 & 14.9 $\pm$ 1.9 & $-12.5$ $\pm$ 1.6 & $-8.1$ $\pm$ 1.0 \\ \hline
			7 & 04-04-2022 & 20:22 & 22:07 & $-30$ & $-17$ & 0.79 & 0.6 $\pm$ 1.6 & 0.6 $\pm$ 1.5 & 0.2 $\pm$ 0.5 \\ \hline
			8 & 25-05-2022 & 17:22 & 18:38 & $-19$ & 41 & 0.81 & 7.1 $\pm$ 1.7 & $-6.8$ $\pm$ 1.6 & 2.2 $\pm$ 0.5 \\ \hline
			9 & 16-07-2022 & 02:31 & 05:00 & 37 & 5 & 1.00 & 15.6 $\pm$ 2 & 13.7 $\pm$ 1.5 & 11.8 $\pm$ 1.3 \\ \hline
			10 & 14-10-2022 & 21:38 & 23:08 & $-25$ & 36 & 0.65 & 3.7 $\pm$ 1.9 & $-2.6$ $\pm$ 1.3 & $-2.9$ $\pm$ 1.4 \\ \hline
			11 & 10-02-2023 & 02:30 & 03:42 & 27 & 24 & 0.31 & 3.7 $\pm$ 3 & 3.5 $\pm$ 2.8 & 1.4 $\pm$ 1.1 \\ \hline
			12 & 24-02-2023 & 19:40 & 21:30 & 32 & 19 & 0.95 & 15.9 $\pm$ 2.1 & 11.9 $\pm$ 1.3 & 14.6 $\pm$ 1.6 \\ \hline
			13 & 25-02-2023 & 18:40 & 20:00 & 27 & 37 & 0.62 & 13.1 $\pm$ 2.1 & 1.7 $\pm$ 0.3 & 11.7 $\pm$ 2.1 \\ \hline
			14 & 09-04-2023 & 17:55 & 20:13 & $-20$ & 43 & 0.79 & 0.7 $\pm$ 1.6 & $-0.4$ $\pm$ 0.9 & $-0.6$ $\pm$ 1.3 \\ \hline
			15 & 17-04-2023 & 11:28 & 13:43 & 10 & $-17$ & 0.74 & 11.4 $\pm$ 2 & $-11.0$ $\pm$ 1.9 & $-3.0$ $\pm$ 0.5 \\ \hline
			16 & 21-04-2023 & 17:44 & 18:53 & $-22$ & 13 & 1.17 & 1.9 $\pm$ 1.3 & $-1.7$ $\pm$ 1.1 & $-1.0$ $\pm$ 0.7 \\ \hline
			17 & 14-12-2023 & 16:45 & 17:24 & 5 & 49 & 0.98 & 17.7 $\pm$ 2.2 & $-8.1$ $\pm$ 1 & 15.8 $\pm$ 2 \\ \hline
			18 & 20-01-2024 & 08:20 & 09:18 & $-16$ & $-24$ & 0.60 & 8.3 $\pm$ 2 & 3.3 $\pm$ 0.8 & $-7.8$ $\pm$ 1.8 \\ \hline
			19 & 10-02-2024 & 22:45 & 23:40 & $-10$ & 13 & 0.40 & 0.5 $\pm$ 2.5 & 0.3 $\pm$ 1.7 & 0.4 $\pm$ 1.8 \\ \hline
			20 & 03-05-2024 & 02:06 & 03:12 & 25 & $-6$ & 0.75 & 13.0 $\pm$ 2.2 & 11.4 $\pm$ 1.7 & $-7.9$ $\pm$ 1.2 \\ \hline
			21 & 28-07-2024 & 00:52 & 02:26 & $-10$ & $-14$ & 0.77 & 12.1 $\pm$ 2 & 0.4 $\pm$ 0.1 & 12.3 $\pm$ 2 \\ \hline
			22 & 22-04-2025 & 06:31 & 08:30 & $-30$ & $-10$ & 0.72 & 13.7 $\pm$ 2.1 & $-13.6$ $\pm$ 2.1 & 1.6 $\pm$ 0.2 \\ \hline
			23 & 30-05-2025 & 23:31 & 00:51 & 14 & $-10$ & 0.36 & 0.4 $\pm$ 2.7 & 0.1 $\pm$ 0.7 & $-0.4$ $\pm$ 2.6 \\ \hline
			24 & 05-08-2025 & 15:37 & 16:17 & 6 & 4 & 0.90 & 20.5 $\pm$ 2.4 & $-12.0$ $\pm$ 1.4 & 16.8 $\pm$ 2 \\ \hline
			25 & 03-10-2025 & 04:48 & 05:43 & 12 & 11 & 0.85 & 14.0 $\pm$ 2 & 5.0 $\pm$ 0.7 & $-13.6$ $\pm$ 1.9 \\ \hline
			26 & 12-10-2025 & 13:04 & 15:22 & 23 & 11 & 1.29 & 3.3 $\pm$ 1.3 & 3.1 $\pm$ 1.2 & 1.2 $\pm$ 0.5 \\ \hline
			27 & 05-11-2025* & 22:00 & 22:33 & 30 & $-42$ & 0.94 & 14.8 $\pm$ 1.9 & $-1.7$ $\pm$ 0.2 & 13.2 $\pm$ 1.9 \\ \hline
			28 & 09-11-2025* & 07:00 & 09:00 & 25 & $-9$ & 0.82 & 12.0 $\pm$ 1.9 & 11.6 $\pm$ 1.8 & $-3.8$ $\pm$ 0.6 \\ \hline
			29 & 11-11-2025* & 09:45 & 10:48 & 24 & 23 & 0.82 & 9.7 $\pm$ 1.8 & $-7.4$ $\pm$ 1.3 & $-6.6$ $\pm$ 1.2 \\ \hline
			30 & 14-11-2025* & 08:30 & 09:42 & 22 & 60 & 0.54 & 1.5 $\pm$ 2.1 & $-1.5$ $\pm$ 2.1 & 0.1 $\pm$ 0.1 \\ \hline
			31 & 18-01-2026 & 18:00 & 18:13 & -9 & -15 & 1.01 & 13.2 $\pm$ 2.1 & $-8.8$ $\pm$ 1.5 & $-12.9$ $\pm$ 1.5 \\ \hline
		\end{tabular}%
	\tablecomments{We list here the event number, date, start and end times of the dimming impulsive phase, heliographic location of the associated flare, and the DIRECD results including the height of the best-fit cone (in $R_{\sun}$), the 3D inclination angle, and the inclination angles projected onto the meridional and equatorial planes.The positive and negative values represent North/West and South/East directions. For events marked with *, we used SDO ``quick-look'' data that is immediately available in near real-time.}
	\label{table:events}
\end{table}

\subsection{Catalog content and data products}
 \label{catalog}
The DIRECD CME catalog provides a standardized set of data products derived consistently for all cataloged events using the DIRECD software. For each event listed in Table~\ref{table:events}, we apply the full analysis pipeline, from coronal dimming identification to CME direction estimation in a uniform and automated manner.

The end-to-end DIRECD processing produces a standardized set of data products for each cataloged event. All products are archived and publicly available through the Zenodo database\footnote{{\url{https://doi.org/10.5281/zenodo.20488339}}}.  The released dataset contains a total of 64 cataloged CMEs spanning 2010-2026. The dataset comprises 31 newly analyzed CMEs that occurred during solar cycle 25 (2021-2026) that are discussed in detail in this study, plus 33 additional events adopted from earlier DIRECD statistical samples (2010-2021). The core output of the DIRECD analysis is the 3D CME propagation direction inferred from coronal dimming geometry at the end of the dimming impulsive phase. The 3D CME propagation direction is specified by the heliographic longitude and latitude of the  source and the 3D coordinates of the cone apex, together with the inclination angle relative to the local radial direction. In addition, the catalog provides the CME angular width and the cone height at which the erupting structure remains connected to the dimming footprint. For completeness, two-dimensional projections of the inclination angle onto the meridional and equatorial planes are also included. Quantitative DIRECD results, including CME cone parameters, and dimming area time series (where available), are provided as plain-text files. Spatial products, including cumulative and timing-resolved coronal dimming masks, are distributed in FITS format. The availability of pre-computed dimming masks allows users to bypass the dimming detection step and proceed directly to CME direction analysis, optionally adjusting model parameters for specific applications and research needs. These products are provided for all 31 newly analyzed Solar Cycle~25 events listed in Table~\ref{table:events}; the public repository also includes the corresponding dimming masks and DIRECD parameters for the additional legacy events adopted from earlier DIRECD statistical samples \citep{jain2025validating}.

These derived geometrical parameters inherently contain uncertainties arising from both the intrinsic challenges of coronal dimming detection and fundamental observational limitations. To rigorously quantify these uncertainties, we developed a statistical framework that propagates the two dominant sources of error through Monte Carlo simulations. We assume that the precision of CME source location on the solar disk carries a conservative range of uncertainty of 20 pixels corresponding to approximately $\pm$ 48 arcseconds given the plate scale of 2.4 arcseconds per pixel for a $1024 \times 1024$ SDO image. Concurrently, the determination of CME cone height exhibits a relative uncertainty of 10\%, translating to approximately 0.1 solar radii, which reflects the limitations in reconstructing precise three-dimensional geometry from two-dimensional projected imagery. We employ comprehensive Monte Carlo simulations with 10,000 iterations to propagate these uncertainties through the DIRECD geometrical model. For each CME event, the simulation incorporated random perturbations within the specified uncertainty ranges for both source location and cone height. The standard deviation of all valid simulations provides our $1\sigma$ uncertainty estimate for each CME's inclination angle. Columns 7-10 of table~\ref{table:events} lists the derived best-fit height, 3D inclination angle and 2D angles in planes obtained from DIRECD method for the events under study. A subset of the catalog events were analyzed using SDO/AIA 'quick-look' data 
products (Table \ref{table:events}, marked with asterisks*), which are made available 
within minutes of acquisition, rather than the standard level 1.5 data products 
that undergo full calibration. Quick-look data are generated by an expedited, automated pipeline designed to prioritize minimal processing time, thereby making solar observations available to the scientific community and space weather forecast centers within minutes of downlink. For the AIA images, quick-look processing applies only the most rudimentary calibrations, such as bias subtraction and a basic flat-field, to facilitate rapid browsing and event detection. Despite these differences, quick-look products retain sufficient quality for dimming detection and CME direction estimation. To quantify this, we reprocessed all four quick-look events in our catalog (events \#27--30; Table~\ref{table:events}) using the standard, fully calibrated SDO/AIA Level~1.5 data products, applying identical detection parameters, thresholds, and impulsive-phase criteria to those used for the quick-look analysis. For each event, we compare the resulting best-fit cone height and inclination angle derived from the two data products. As evident from Table \ref{table:qlvsl15}, the quick-look and Level~1.5 results agree to within $0.08\,R_\odot$ in cone height (mean difference $0.03\,R_\odot$) and $1.8^{\circ}$ in inclination angle (mean difference $0.85^{\circ}$) across the four events, with the largest discrepancies occurring for event~\#30 in the quick-look subset. In all four cases the differences remain within, or comparable to, the $1\sigma$ Monte Carlo uncertainties reported for each event in Table~\ref{table:events}, confirming that the rudimentary calibration applied to quick-look data does not measurably degrade the DIRECD-derived CME geometry, supporting its use for near-real-time, operational CME direction estimation.

\begin{table}[htbp]
	\centering
	\raggedright
	\small
	\caption{Comparison of DIRECD results using quick-look vs.\ Level~1.5 SDO/AIA data for the four quick-look events}
		\begin{tabular}{|c|l|l|c|c|c|c|}
			\hline
			\textbf{\#} & \textbf{Event Date} & \textbf{Data} & \textbf{Height (H)} & \textbf{Inclination (I)}  & $\boldsymbol{\Delta}$\textbf{(QL$-$L1.5)} \\
			 & & & ($R_{\sun}$) & ($^\circ$) & \\ \hline
			\multirow{2}{*}{27} & \multirow{2}{*}{05-11-2025} & QL   & 0.94 & 14.8  & \multirow{2}{*}{H: 0.01, I: 1.0, } \\
			 & & L1.5 & 0.95 & 13.8  & \\ \hline
			\multirow{2}{*}{28} & \multirow{2}{*}{09-11-2025} & QL   & 0.82 & 12.0 & \multirow{2}{*}{H: 0, I: 0, } \\
			 & & L1.5 & 0.82 & 12.0 & \\ \hline
			\multirow{2}{*}{29} & \multirow{2}{*}{11-11-2025} & QL   & 0.82 & 9.7 & \multirow{2}{*}{H: 0.01, I: 0.6, } \\
			 & & L1.5 & 0.81 & 9.1 & \\ \hline
			\multirow{2}{*}{30} & \multirow{2}{*}{14-11-2025} & QL   & 0.54 & 1.5 & \multirow{2}{*}{H: 0.08, I: -1.8, } \\
			 & & L1.5 & 0.46 & 3.3 & \\ \hline
		\end{tabular}
	\tablecomments{QL = quick-look; L1.5 = standard, fully calibrated SDO/AIA Level~1.5 data. Event numbering follows Table~\ref{table:events}. $\Delta$(QL$-$L1.5) denotes the absolute difference in height (H) and inclination (I) between the two data products for each event.}
	\label{table:qlvsl15}
\end{table}

A particularly strong independent validation of the DIRECD method is provided for the March 28, 2022 event (event \#5), for which a detailed multi-spacecraft case study was performed by \cite{podladchikova2024three}. In that study, the DIRECD method was applied to coronal dimming detected in STEREO-A/EUVI 195 \AA~observations, yielding a best-fit cone height of 1.12 $R_{\sun}$, a full angular width of $42^\circ$, and a CME propagation direction inclined $6^\circ$ from the radial direction. The present catalog derives equivalent parameters for the same event using an entirely independent instrument and passband - SDO/AIA 211 \AA - obtaining a cone height of 1.1 $R_{\sun}$, a full angular width of $45.8^\circ$, and an inclination angle of $2.5^\circ$. The angular separation between the SDO and STEREO-A was $33^\circ$. The two results show a difference of only 0.02 $R_{\sun}$ ($\leq$ 2\%), obtained from two different spacecraft, different viewing geometries, and independent detection pipelines. The angular width likewise agrees to within $\sim$ $4^\circ$, well within the combined uncertainties of the method. The small residual difference in inclination angle ($6^\circ$ vs. $2.5^\circ$) and filament directions ($9.1^\circ$ vs 11 $^\circ$) is within reasonable limits. This agreement across instruments confirms that the DIRECD-derived cone geometry is robust and not an artifact of a particular passband, threshold choice, or viewing angle. The convergence of results across two independent observational chains for this event therefore constitutes a strong available cross-validation of the DIRECD methodology.
	


\begin{figure}[h]
	\centering
	\subfloat{%
		\includegraphics[width=0.99\textwidth]{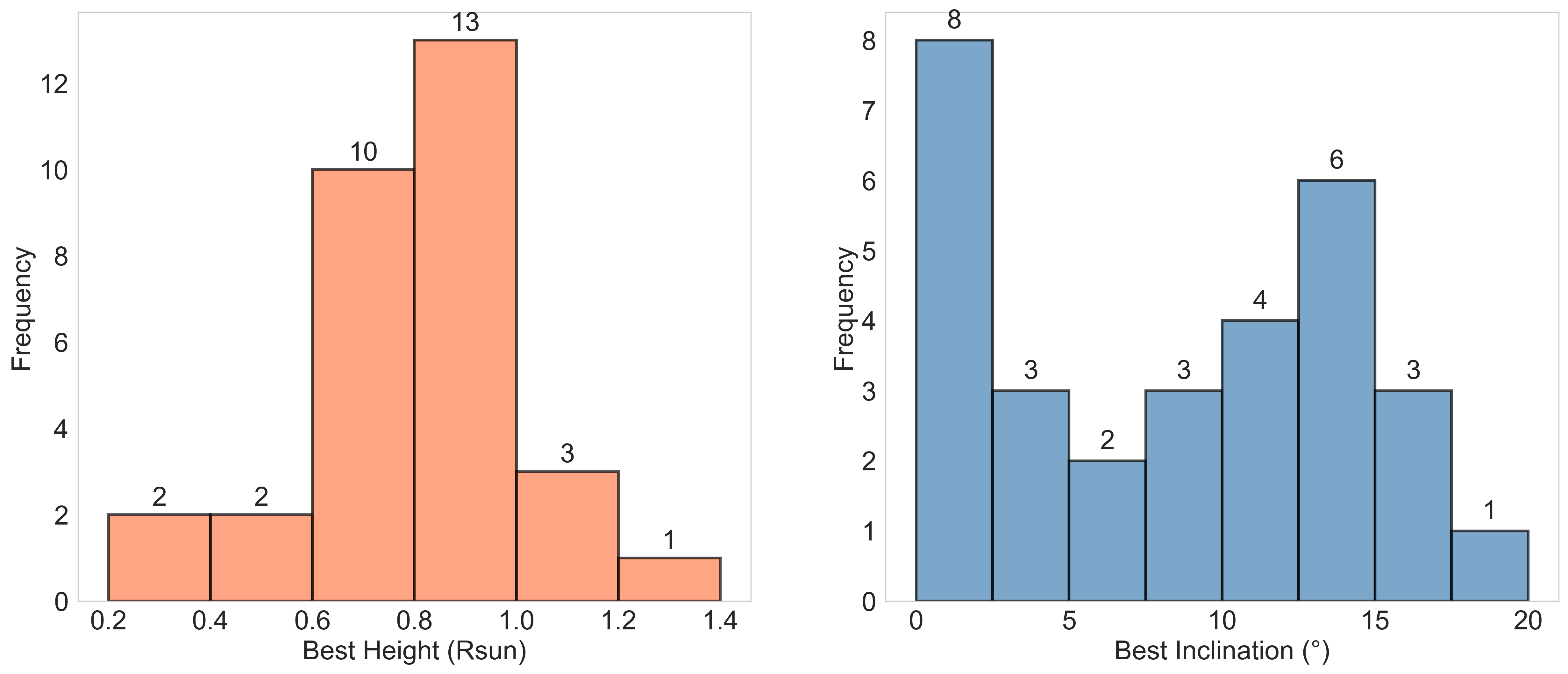}}\qquad
	{\includegraphics[width=0.99\columnwidth]{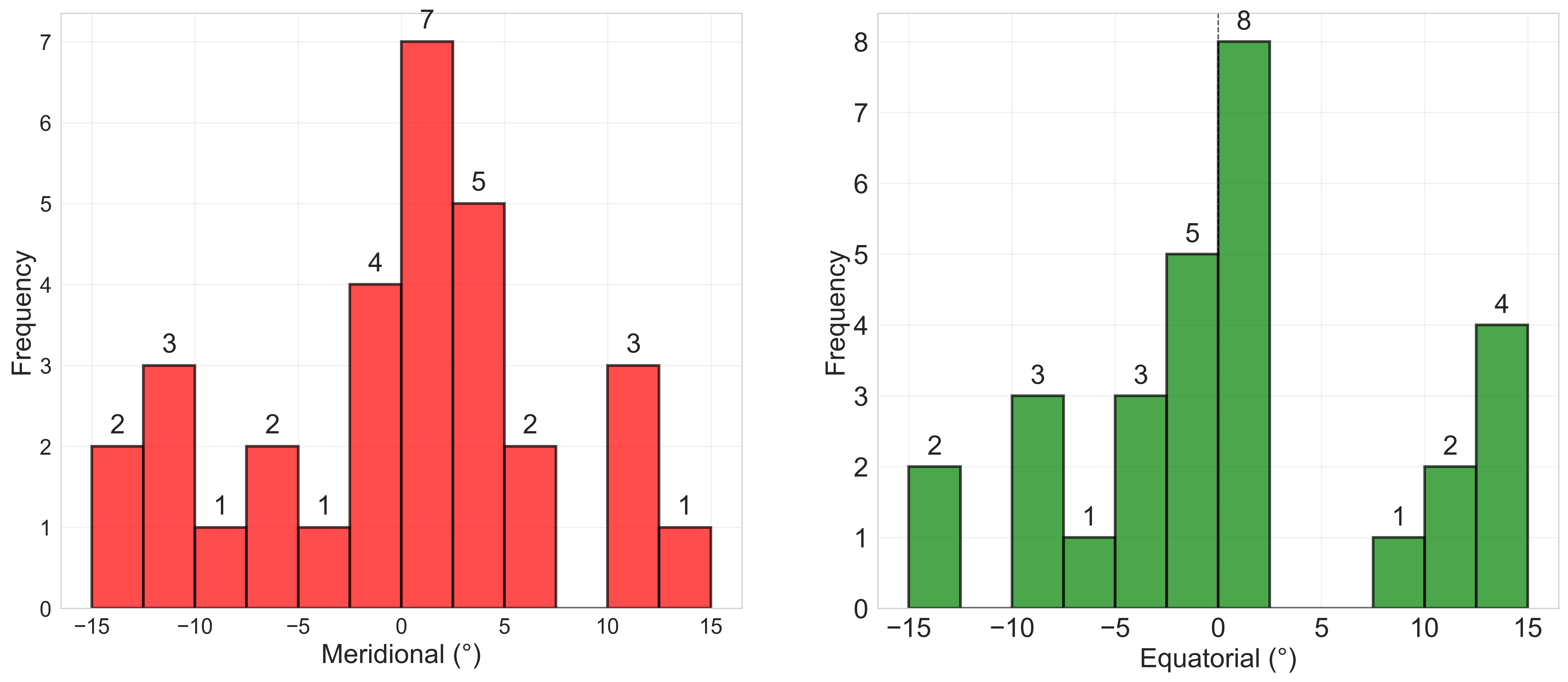}%
	}
	\caption{Histogram distributions of the DIRECD-derived CME geometric parameters for the 31 events in our catalog. From left to right: (a) best-fit cone height (in $R_\sun$); (b) best-fit inclination angle (in degrees); (c) meridional deflection angle (in degrees); and (d) equatorial deflection angle (in degrees).}
	\label{fig:histogram_table}
\end{figure}

Figure~\ref{fig:histogram_table} presents the distributions of heights, inclination angles in 3D and angles in plane derived from coronal dimming geometry using DIRECD. The histogram of inclination angles reveals that approximately half of the CMEs (16/31 events) have low inclination angles ($\leq 10^\circ$), while another 15 events exhibit moderate inclinations in the $10–20^\circ$ range. The best-fit cone height distribution shows a pronounced concentration of events between 0.6 and $1.0 R_{\sun}$, comprising 23 of 31 CMEs (74\%) in our catalog. This finding has significant implications for understanding CME initiation: This height range corresponds to the altitude to which the dimming signatures 
extend in our observations. The DIRECD cone height is determined by the height 
at which the dimming area growth rate stabilizes; above this height, the dimming 
geometry no longer constrains the cone geometry. This does not imply loss of magnetic connectivity-
the CME remains connected to the solar surface at all heights-but rather reflects 
the limit of geometric constraints available from coronal dimming observations. To further characterize the 3D propagation directions, we show the histograms of meridional and equatorial
components. The histograms of the projections of the 3D inclination into the meridional (longitidudinal) and equatorial (latitudinal) planes reveals that half of the events have a low deflection of $\pm 5^\circ$ in longitudinal and latitudinal direction. 

\begin{figure}
	\centering
	\includegraphics[width=\textwidth]{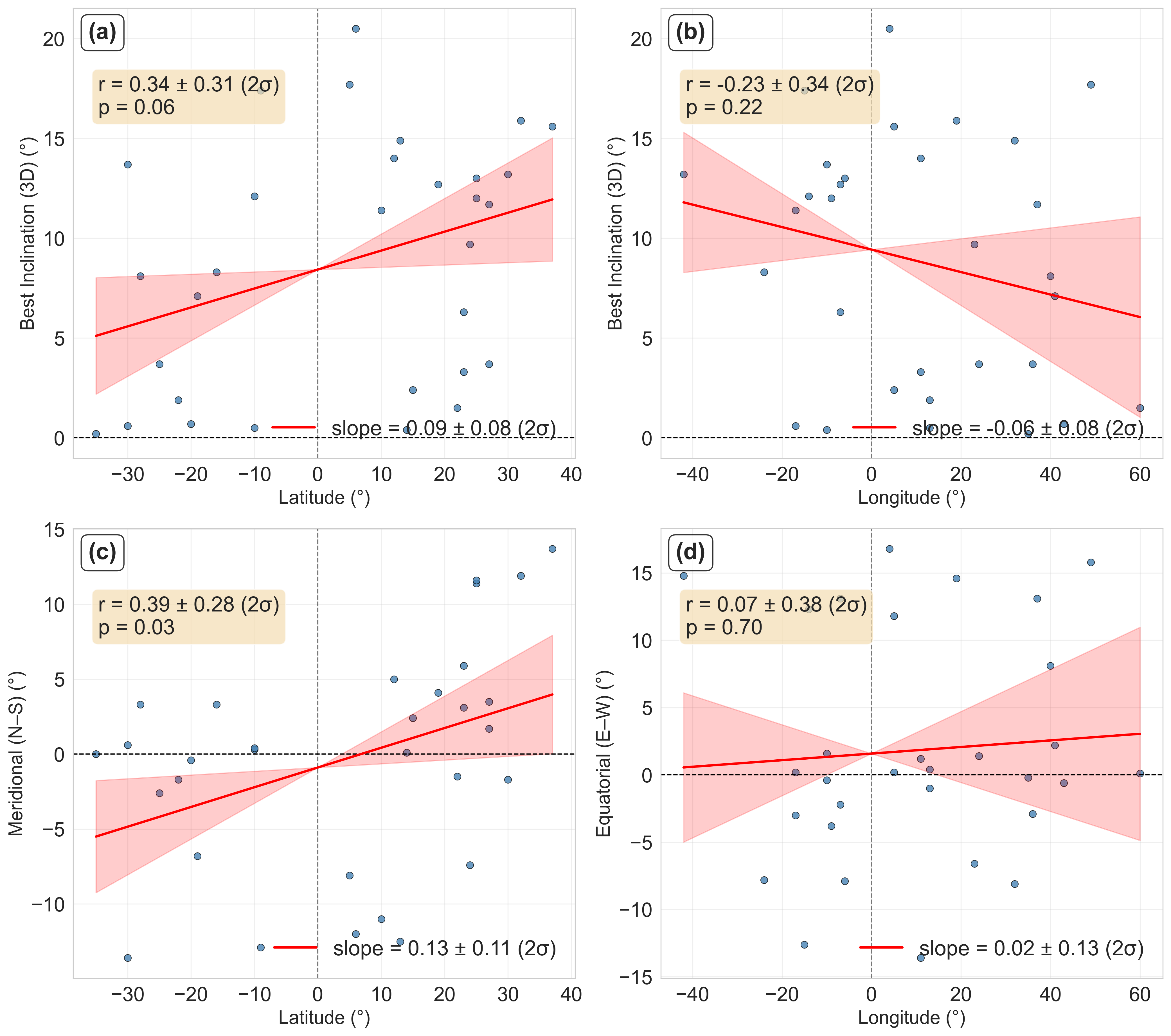}
	\caption{Scatter plots of CME inclination components versus heliographic position of the source active region. Panel~(a): best-fit 3D inclination as a function of heliographic latitude; panel~(b): best-fit 3D inclination as a function of heliographic longitude; panel~(c): meridional (N–S) inclination component as a function of heliographic latitude; panel~(d): equatorial (E–W) inclination component as a function of heliographic longitude. The red line shows a linear least-squares fit, with the slope $\pm 2\sigma$ indicated in each panel. The shaded region shows the $\pm 2\sigma$ bootstrap confidence band on the fit. The correlation coefficient r$\pm 2\sigma$ and associated p-value, both derived from bootstrapping, are given in the upper corner of each panel.}
	\label{fig:summary_table}
\end{figure}

Figure~\ref{fig:summary_table} displays the heliographic distribution of source locations as a function of heliographic longitude and latitude, for each of the three inclination measures. The inclination angles vary between $0-20^\circ$ while the heights are located within the low corona (typically $\leq$\(2 R_{\sun}\)). The spatial spread across longitudes and latitudes reflects the broad distribution of source active regions, without strong preferential latitudinal 
bias or longitudinal bias.

Panels~(a) and~(b) show the total 3D inclination against latitude and longitude respectively. Panel (a) shows that there is a weak positive trend (slope = $0.09 \pm 0.08$, r= $0.34 \pm 0.31$, p=0.06) which is close to the 5\%
significance level. Against longitude (panel b), no trend is found (r = $-0.23 \pm 0.34$, p = $0.22$). The more informative picture comes from separating the inclination into its meridional and equatorial components. Panel~(c) shows the meridional component against latitude. The positive slope ($0.13 \pm 0.11$, r = $0.39\pm 0.28$, p = $0.03$) indicates that events from the northern hemisphere reveal a tendency to be deflected northward, while those from the southern hemisphere tend to be deflected southward. As a robustness check, Appendix \ref{appendixA} repeats this analysis including the 33 legacy events adopted from earlier DIRECD statistical samples. Panel~(d) of figure \ref{fig:summary_table} shows that the equatorial component does not show any strong dependence on longitude (slope = $0.02 \pm 0.13$, r = $0.07 \pm 0.38$, p = 0.70) and the values are distributed symmetrically about zero across the full spatial range sampled. This east–west symmetry is consistent with the large-scale axial symmetry of the coronal magnetic field about the solar rotation axis, and serves as an internal consistency check: unlike the meridional component, which has a geometrically preferred direction (the equator), there is no equivalent large-scale structure that would systematically bias CME propagation eastward or westward. Furthermore, no clear relationship was found between CME speed and longitudinal deflection, suggesting that east-west propagation in the low corona is not strongly controlled by eruption kinematics, but is instead dominated by local and likely stochastic variations in the surrounding magnetic environment. 

These results indicate a systematic tendency for early CME propagation to occur away from the heliographic equator, in contrast with the equatorward deflection commonly observed at larger heliocentric distances, where global magnetic pressure gradients associated with the heliospheric current sheet (HCS) redirect trajectories toward the solar equatorial plane \citep{cremades2004, kay2015heliocentric}. Within the low corona ($\lesssim 2R_{\sun}$), the initial trajectory is therefore governed primarily by local active-region magnetic field geometry, with erupting flux ropes expelled preferentially along paths of lower magnetic confinement away from the dense, strongly magnetised core of the source region. This local control is nevertheless modulated by the large-scale coronal environment: low-latitude source regions sit closer to the HCS, where the equatorward pressure gradient of the global field may partially offset the poleward tendency, whereas high-latitude regions erupt into an environment where that gradient is weaker and local field geometry dominates more completely. The latitude dependence of deflection within each hemisphere thus suggests that the transition from local active-region control to global coronal-field guidance may already begin to manifest within the low corona, before becoming the dominant influence at greater heights. Southern-hemisphere statistics are too sparse for a comparable latitude-dependent analysis, though the available events are predominantly poleward, consistent with the overall hemispheric trend. The longitude analysis offers no analogous structure: whether events originate from eastern or western longitudes has no bearing on the equatorial deflection component, confirming the absence of any systematic east–west geographic dependence.

The catalog is designed to be reusable and extensible. In addition to the 31 newly analyzed Solar Cycle 25 events presented in Table \ref{table:events}, the publicly released dataset incorporates CME events from earlier statistical samples \citep{jain2025validating} analyzed using the same DIRECD methodology. For these events, the corresponding coronal dimming masks and dimming-derived CME direction parameters are included in the online catalog, forming a unified dataset spanning multiple event samples.


\subsection{CME Geoeffectiveness and Geomagnetic Storm Prediction}

The prediction of geomagnetic storm intensity remains one of the most pressing 
challenges in operational space weather forecasting. Current 
operational forecasts rely almost exclusively on coronagraph-based observations when CMEs reach heights of 3--5~R$_{\odot}$. This temporal delay critically constrains early warning capability and 
limits the window for hazard mitigation. Here we demonstrate that integration of 
early directional information can help  with geoeffectiveness forecasting, one that operates in near real-time and 
explicitly accounts for whether a CME will actually impact Earth's magnetosphere.

The three-dimensional CME propagation directions determined from low-coronal dimming observations can be leveraged to improve space weather forecasting, particularly the prediction of geomagnetic storm intensity. 
 To assess the practical utility of the DIRECD-derived directions, we computed the correlation between observed geomagnetic disturbance indices from OMNI hourly data \citep{papitashvili2020omni}  and a composite measure combining the measure of early CME direction and reliable CME speed obtained from CDAW/LASCO catalog \citep{gopalswamy2009soho} for 28/31 events. For the remaining 3 events, the CME was either classified as a "poor-event" or its speed was not available in the catalog.

To quantify CME direction, we use the angular miss distance $E$, defined as the angular 
separation between the CME's inferred propagation direction axis (determined from DIRECD analysis of coronal 
dimming geometry) and the Sun-Earth line, approximated using the CME's launch location. Here, Lat and Lon denote the heliographic latitude and longitude of the CME launch site, which is estimated by the flare source region for on-disk events (within $\pm$ $60^\circ$–$75^\circ$). If $\gamma$ is the meridional angle and $\delta$ is the equatorial angle obtained from DIRECD, then the angular miss distance E is given by:

\begin{equation}
	E = \sqrt{(\textrm{Lat} - |\textrm{$\gamma$}|)^2 + (\textrm{Lon} - |\textrm{$\delta$}|)^2}
	\label{eq:miss_distance}
\end{equation}

For each CME in the sample, the associated geomagnetic response was identified through a multi-step temporal association procedure. An expected arrival window at Earth was first estimated from the CME launch time and the plane-of-sky speed $V_{\mathrm{POS}}$ measured from LASCO coronagraph imagery, using a simple ballistic propagation model that assumes
constant velocity over the Sun--Earth distance of 1~AU:
\begin{equation}
	t_{\mathrm{arrival}} = t_{\mathrm{launch}} + \frac{1\,\mathrm{AU}}{V_{\mathrm{POS}}}.
	\label{eq:arrival}
\end{equation}
Within this window, the OMNI solar wind dataset \citep{papitashvili2020omni} was inspected for signatures consistent with an interplanetary CME (ICME) or driven shock arrival: a sustained enhancement in solar wind speed, an increase in magnetic field magnitude (Bz), a rise in proton density, and, where present, a coherent rotation in the magnetic field vector indicative of a magnetic cloud. For each CME, we identified the minimum Dst value within an expected CME arrival window.

\begin{figure}[h]
	\centering
	\includegraphics[width=\textwidth]{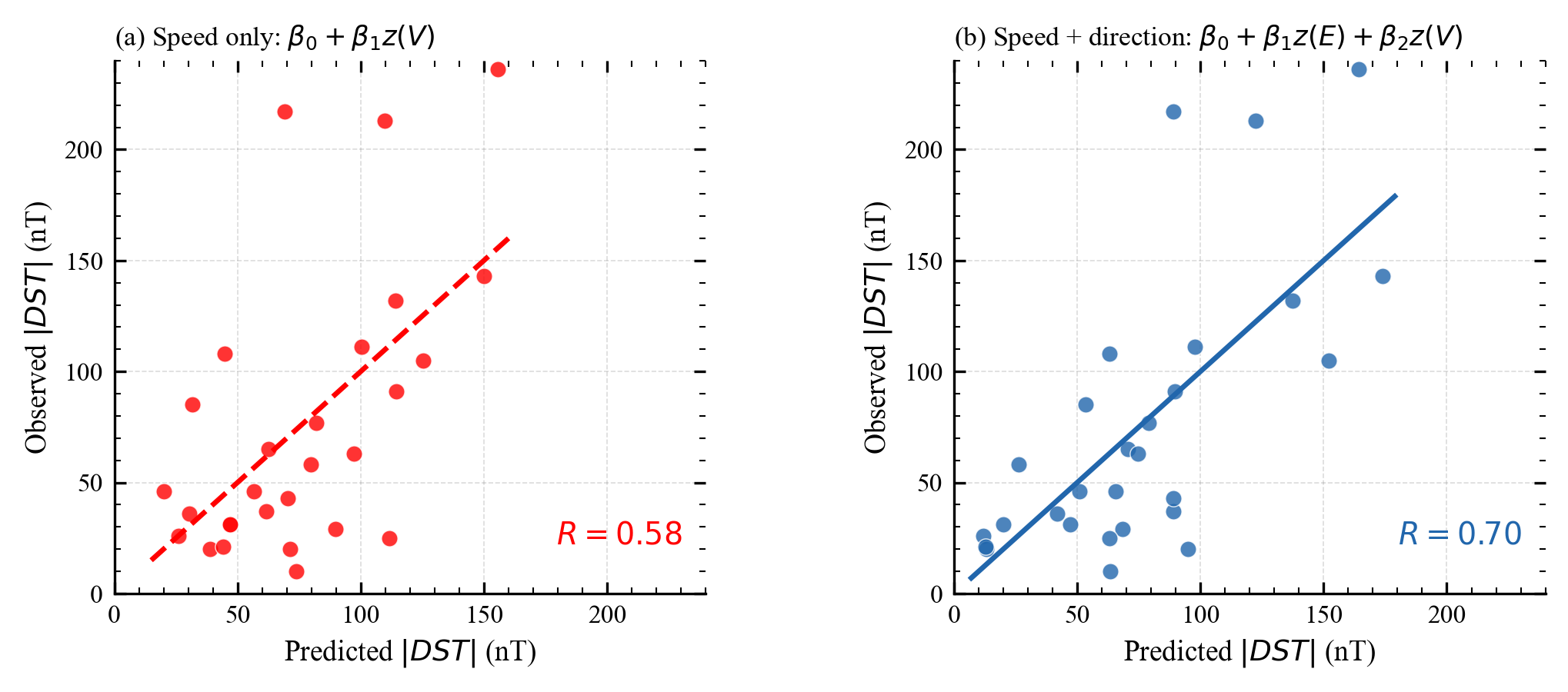}
	\caption{Observed versus predicted geomagnetic storm intensity for 25 CME events. Left panels shows baseline model which uses only the CME speed, the right panel shows the combined model incorporating both CME speed and propagation direction.  The colored line shows the linear regression fit. Adding the DIRECD-derived directional term improves the correlation from R=0.58 to R=0.7.}
	\label{fig:geoeffectiveness}
\end{figure}

We acknowledge that for events producing only weak geomagnetic responses (Dst above approximately $-30$~nT), a unique one-to-one attribution between a specific CME and the observed Dst perturbation is inherently ambiguous: the disturbance could equally originate from a co-rotating interaction region
(CIR), a high-speed stream (HSS), or a different, temporally proximate CME. Furthermore, not every CME in the catalog is expected to impact Earth's magnetosphere; a CME may miss Earth entirely or produce no discernible geomagnetic signature. For closely spaced or interacting CMEs, the assigned Dst values represent the associated geomagnetic response within the arrival window and do not imply a unique physical attribution. Nevertheless, we retain all events, including
non-storm cases, in the statistical sample to avoid selection bias toward geoeffective events and to sample the full spectrum of CME--geomagnetic coupling, from negligible disturbances to intense storms. The Dst values listed in Table~\ref{tab:cme_angular_distance} should therefore be
interpreted as the geomagnetic activity level temporally associated with each CME, rather than as a definitive causal attribution in every individual case.

\begin{equation}
	|Dst| = \beta_0 + \beta_1 \, z(E) + \beta_2 \, z(V)
	\label{eq:geoeff}
\end{equation}
where $|Dst|$ is the absolute magnitude of the Dst geomagnetic disturbance index. Since the predictors are in different units, we standardize them as $z(\cdot) = (x - \mu_x) / \sigma_x$ to ensure fair relative weighting, preventing the speed variable (which spans 200--1800~km~s$^{-1}$) from dominating the geometric term (which spans $\sim$10--60 degrees). Linear regression of Equation~\eqref{eq:geoeff} on the 28 events with complete data results to a  of  $r = 0.7$, indicating that the model has a moderate-high correlation with the geomagnetic storm intensity. The resulting best-fit regression coefficients (n = 28) are $\beta_{0}$ = 75.9 $\pm$ 8.8 nT, $\beta_{1}$ = -24.3 $\pm$ 9.1, and $\beta_{2}$ = 41.5 $\pm$ 9.1, where uncertainties denote 1$\sigma$ standard errors. We note that these coefficients apply only to inputs standardized using the sample mean and standard deviation of E and V; they are not intended for direct application to raw, unstandardized values. This is a substantial improvement over speed alone, which achieves $r = 0.58$ (best-fit regression coefficients: $\beta_{0}$ = 75.9 $\pm$ 9.8 nT and $\beta_{1}$ = 36.1 $\pm$ 9.8) as shown in Figure \ref{fig:geoeffectiveness}. The adjusted R$^2$ increases from 0.32 for the speed-only model to 0.45 once the angular miss distance term is included, indicating that the improvement is not simply an artifact of the additional free parameter. It must be noted that the geo-effectiveness also critically depends on the CME's magnetic field strength and orientation at 1 AU, in particular the duration of southward interplanetary magnetic field (IMF) \citep{kilpua2017coronal, Podladchikova2012, Podladchikova2018}. While $E$ captures the geometric encounter probability, it does not incorporate the CME's internal magnetic structure. 

Following the standard classification \citep{gonzalez1994geomagnetic}, the sample comprises 7 non-storm events ($|Dst|$ $\leq 30$ nT), 7 weak storms ($30~nT \leq~|Dst|~\leq 50$ nT), 6 moderate storms ($50~nT \leq~|Dst|~\leq 100$ nT), and 8 intense storms ($|Dst|$ $\geq 100$ nT), with 3 events lacking speed measurements excluded from the classification. All eight intense storms are associated with angular distances below 35$^\circ$, and six of the eight have angular distances below 25$^\circ$. Conversely, no event with an angular miss distance exceeding 45$^\circ$ produces a storm of moderate or greater intensity, regardless of CME speed: events \#27 and \#30, with speeds of 920 and 1308 km~s$^{-1}$ respectively, yield Dst minima of only -58 and -25 nT owing to their angular miss distances of 63.5$^\circ$ and 63.3$^\circ$. This can be attributed to the geometrical interpretation that at large angular offsets Earth intercepts only the magnetic flank of the ejecta, where the field strength and southward component are substantially reduced relative to the flux-rope core \citep{rudisser2024understanding, rodriguez2020clustering,feng2006geoeffective,zhao2005tentative}. 

Among events with angular distances below 25$^\circ$, the four events exceeding 1300 km~s$^{-1}$ (\#3, \#12, \#29, \#31) all produce intense storms with Dst $\leq$ -105 nT, whereas slower events at comparable angular distances produce weak to moderate responses. However, speed alone is insufficient, as demonstrated by events \#27 and \#30, that even fast CMEs fail to produce significant geomagnetic activity when the propagation direction deviates substantially from the Sun–Earth line. The joint condition of angular miss distance below approximately 35$^\circ$ and plane-of-sky speed above approximately 500 km~s$^{-1}$ correctly identifies the moderate-to-intense storm events in all but one case (\#23, 496 km~s$^{-1}$, 17.4$^\circ$, Dst = $-108$ nT), which marginally violates the speed threshold yet produces an intense storm, likely reflecting a particularly favourable magnetic field orientation at Earth for that event. Across the moderate storm class, the observed vs predicted Dst agreement is closest, with residuals generally within 15 nT. For the seven non-storm events, the model consistently overpredicts the geomagnetic response, in several cases by large margins, for example events \#18, \#19, \#22, and \#30 are predicted to produce moderate storms of -63,
-95,-68, and -63 nT respectively, yet the observed Dst minima remain below -30 nT in all four cases. As noted above, these events are characterised by angular miss distances exceeding 37$^\circ$, and the overprediction at large angular offsets suggests that the angular miss distance term in the regression, while necessary, does not fully capture the nonlinear attenuation of geomagnetic impact at high miss angles. A CME flank encounter not only reduces the peak field strength but also shortens the duration of southward field exposure at Earth, both effects acting to suppress ring-current injection below what a linear angular correction would predict \citep{rudisser2024understanding, rodriguez2020clustering,feng2006geoeffective,zhao2005tentative}.

At the opposite extreme, the model underpredicts the most intense storms in the sample although still classifying as extreme-storm events. Events \#16, \#28, and \#31 all exceed their predictions by 50 nT or more while still being in the severe storm category. This may be because of the several reasons: first, the plane-of-sky speeds used here are projection-corrected only approximately, and for events propagating close to the Sun–Earth line the true radial speed may substantially exceed the LASCO measurement; second, exceptionally large and sustained southward $B_{z}$ excursions, which are not predictable from coronagraph data alone, can drive ring-current injection far beyond what speed-based empirical relations anticipate \citep{richardson2011geoeffectiveness}; third, events \#28 and \#29 occur within 48 hours of one another and may represent a compound or cannibalism scenario in which the preceding ejecta preconditions the magnetosphere, lowering the threshold for ring-current injection and amplifying the Dst response of the trailing event \citep{rodriguez2020clustering}. The underprediction of extreme storms is a known and widely reported limitation of empirical Dst forecasters \citep{kim2010empirical} and further shows that the physics-based modelling, incorporating in-situ magnetic field data, is necessary to supplement the early-warning capability demonstrated here. The early CME propagation direction is the primary binary classifier of geoeffectiveness, while speed modulates intensity for well-directed events, and residual variance is attributable to interplanetary magnetic field evolution that is inaccessible to near-Sun remote sensing. 


\begin{table}[]
	\raggedright
	\centering
	\caption{CME events with speed, Dst index, and angular miss distance between source location and deflection plane.}
	\label{tab:cme_angular_distance}
	\small
		\begin{tabular}{|c|c|c|c|c|c|c|c|}
			\hline
			\textbf{\#} & \textbf{Date} & \textbf{Start Time} & \textbf{Speed} & \textbf{Angular} & \textbf{Dst}  & \textbf{Observed Dst} & \textbf{Predicted Dst} \\
			& & & \textbf{(km/s)} & \textbf{Distance (°)} &\textbf{Peak Time} &\textbf{(nT)} & \textbf{(nT)} \\
			\hline
			1  & 28-09-2021 & 05:23 & 524  & 44.7 & 01-10-2021/14:00 & $-$31  & -20  \\ \hline
			2  & 09-10-2021 & 06:20 & 712  & 25.0 & 12-10-2021/14:00 & $-$65  & -70  \\ \hline
			3  & 02-11-2021 & 01:20 & 1473 & 19.4 & 04-11-2021/13:00 & $-$105 & -152 \\ \hline
			4  & 14-01-2022 & 12:17 & 1341 & 49.3 & 14-01-2022/22:00 & $-$91  & -90  \\ \hline
			5  & 28-03-2022 & 10:46 & 702  & 13.5 & 31-03-2022/17:00 & $-$37  & -89  \\ \hline
			6  & 30-03-2022 & 17:21 & 641  & 23.9 & 01-04-2022/06:00 & $-$46  & -65  \\ \hline
			7  & 04-04-2022 & 20:22 & 271  & 35.1 & 08-04-2022/00:00 & $-$26  & -12  \\ \hline
			8  & 25-05-2022 & 17:22 & 1134 & 46.7 & 28-05-2022/07:00 & $-$63  & -75  \\ \hline
			9  & 16-07-2022 & 02:31 & ---  & 24.2 & -                & -      & -    \\ \hline
			10 & 14-10-2022 & 21:38 & 424  & 43.1 & 17-10-2022/03:00 & $-$20  & -14  \\ \hline
			11 & 10-02-2023 & 02:30 & ---  & 32.6 & -                & -      & -    \\ \hline
			12 & 24-02-2023 & 19:40 & 1336 & 20.6 & 27-02-2023/12:00 & $-$132 & -137 \\ \hline
			13 & 25-02-2023 & 18:40 & 1170 & 34.8 & 27-02-2023/21:00 & $-$111 & -98  \\ \hline
			14 & 09-04-2023 & 17:55 & 490  & 47.1 & 15-04-2023/10:00 & $-$21  & -12  \\ \hline
			15 & 17-04-2023 & 11:28 & 322  & 20.0 & 19-04-2023/14:00 & $-$36  & -42  \\ \hline
			16 & 21-04-2023 & 17:44 & 1284 & 26.5 & 24-04-2023/05:00 & $-$213 & -122 \\ \hline
			17 & 14-12-2023 & 16:45 & 948  & 33.4 & 17-12-2023/16:00 & $-$77  & -79  \\ \hline
			18 & 20-01-2024 & 08:20 & 850  & 37.2 & 24-01-2024/04:00 & $-$15  & -63  \\ \hline
			19 & 10-02-2024 & 22:45 & 817  & 16.3 & 14-02-2024/03:00 & $-$20  & -95  \\ \hline
			20 & 03-05-2024 & 02:06 & 808  & 19.4 & 06-05-2024/01:00 & $-$43  & -89  \\ \hline
			21 & 28-07-2024 & 00:52 & 522  & 28.2 & 30-07-2024/11:00 & $-$31  & -47  \\ \hline
			22 & 22-04-2025 & 06:31 & 1042 & 45.2 & 24-04-2025/20:00 & $-$29  & -68  \\\hline
			23 & 30-05-2025 & 23:31 & 496  & 17.4 & 03-06-2025/10:00 & $-$108 & -63  \\\hline
			24 & 05-08-2025 & 15:37 & 338  & 14.1 & 09-08-2025/10:00 & $-$85  & -53  \\\hline
			25 & 03-10-2025 & 04:48 & 196  & 7.5  & 07-10-2025/12:00 & $-$46  & -51  \\\hline
			26 & 12-10-2025 & 13:04 & ---  & 22.2 & -                & -      & -    \\\hline
			27 & 05-11-2025 & 22:00 & 920  & 63.5 & 08-11-2025/07:00 & $-$58  & -26  \\\hline
			28 & 09-11-2025 & 07:00 & 792  & 18.5 & 11-11-2025/04:00 & $-$217 & -89  \\\hline
			29 & 11-11-2025 & 09:45 & 1772 & 23.4 & 13-11-2025/07:00 & $-$143 & -174 \\\hline
			30 & 14-11-2025 & 08:30 & 1308 & 63.3 & 16-11-2025/23:00 & $-$25  & -63  \\\hline
			31 & 18-01-2026 & 18:00 & 1842 & 33.1 & 20-01-2026/16:00 & $-$236 & -165\\
			\hline
		\end{tabular}%
			\tablecomments{We list here the event number, date, start and end times of the CME/dimming event analysis, CME speed, Angular miss distance, Dst Peak time, observed Dst from OMNIWeb database\footnote{\url{https://omniweb.gsfc.nasa.gov/form/dx1.html}} and predicted Dst from the regression. }
\end{table}

\section{Discussion and Conclusion}\label{conclusions}
This paper presents a CME direction catalog that combines 31 newly analyzed Solar Cycle~25 events with 33 events adopted from earlier DIRECD statistical samples, resulting in a total of 64 cataloged CMEs and providing a validated, extensible reference dataset for early CME propagation studies. The catalog provides a standardized and reusable set of dimming-derived CME propagation parameters in the low corona, including the three-dimensional direction, inclination in the meridional and equatorial plane (lat/lon), angular width, and the cone height at which the erupting structure remains connected to the dimming footprint, together with coronal dimming masks, timing information, and dimming area evolution products used to derive these parameters. The DIRECD software successfully operationalizes the DIRECD method, offering a user-friendly, web-based interface that automates the workflow from SDO/AIA data acquisition and calibration to dimming detection and 3D CME cone modeling.

Using two representative events from the catalog (24 February 2023 and 18 January 2026), we demonstrated that the DIRECD workflow can robustly reconstruct CME cone geometries directly from the dimming footprint at the end of the dimming impulsive phase, i.e., at heights well below the coronagraph occulting disk where traditional white-light reconstructions are unavailable. The resulting catalog distributions show that the majority of events exhibit low-to-moderate inclination angles, ranging from $0.2^\circ$ to $20.5^\circ$, with 16 of 31 events (52\%) having inclinations $\leq 10^\circ$ and 14 events (45\%) falling in the $10^\circ$--$20^\circ$ range. The best-fit cone heights span $0.31$--$1.29\,R_{\sun}$ and cluster strongly in the low corona, with 23 of 31 events (74\%) concentrated between $0.6$ and $1.0\,R_{\sun}$. The catalog supports quantitative downstream use in CME modeling, inter-comparison studies, and space weather forecasting applications by providing the best-fit CME geometry parameters and associated $1\sigma$ uncertainties derived from Monte Carlo propagation of the dominant error sources.

The decomposition of inclination angles into meridional and equatorial components reveals a clear asymmetry in early CME propagation behaviour. Meridional inclinations exhibit a systematic tendency toward latitude-dependent deflection, with poleward deflection being more pronounced at higher source latitudes, whereas equatorial inclinations remain symmetrically distributed about zero with no significant dependence on longitude. This indicates that the early propagation direction of CMEs is influenced by source latitude, but not by source longitude. The observed poleward behaviour for CMEs occurring at high latitudes, in contrast to the equatorward deflections reported from coronagraph observations at larger heliocentric distances, provides observational evidence of a distance-dependent CME deflection. The results suggest that the earliest stages of CME propagation are governed primarily by the local magnetic field geometry of the source active region, while the influence of the large-scale coronal magnetic field and the heliospheric current sheet becomes increasingly important at greater heights. The tendency for lower-latitude events to exhibit weaker poleward deflections further suggests that this transition between local and global magnetic control may already begin within the low corona for eruptions occurring close to the solar equatorial plane. This highlights the diagnostic value of dimming-based direction estimates for probing the earliest phase of CME propagation. Since DIRECD derives its directional estimates from low-coronal EUV observations during the impulsive phase of the eruption, these constraints become available well before the CME enters the coronagraph field of view, offering a complementary pathway for early CME-driven geomagnetic storm forecasting analogous to what coronal hole observations provide for CIR/HSS-driven storms \citep{Nitti2023}.

Independent cross-validation for March 28, 2022 event against STEREO and SDO spacecrafts further supports the reliability of the dimming-inferred propagation directions and shows strong morphological consistency between the dimming-derived direction and the observed CME expansion axis at larger heliocentric distances. In this sense, the DIRECD catalog provides a bridge between low-coronal signatures and the height ranges commonly used in heliospheric propagation modeling.

We further demonstrate the practical utility of DIRECD-derived directional constraints for space weather forecasting. By constructing a linear geoeffectiveness model that combines the standardized CME speed and the angular miss distance E, a geometric measure of how closely the CME propagation axis aligns with the Sun–Earth line, derived from coronal dimming geometry - we show that incorporating early directional information improves the correlation with observed geomagnetic storm intensity, with correlation increasing from $r=0.58$ to $R=0.7$ when directional information is added to CME speed. Moreover, the regression correctly classifies the moderate-to-intense storm events with the joint condition of angular miss distance below approximately 35$^\circ$ and CME speed above 500 km~s$^{-1}$  in all but one case in our study. Since DIRECD derives its directional estimates from low-coronal EUV observations during the impulsive phase of the eruption, these constraints become available well before the CME enters the coronagraph field of view, offering a meaningful extension of the early warning window for operational space weather forecasting.

An important distinction of the present analysis is that we use the initial plane-of-sky CME speed measured close to the Sun from the CDAW/LASCO catalog \citep{gopalswamy2009soho}, rather than the in-situ ICME speed measured at 1~AU that is commonly employed in CME–Dst correlation studies \citep{srivastava2004solar, yurchyshyn2004correlation, kim2010empirical}. This choice is deliberate and motivated by the forecasting context of this work: the DIRECD-derived propagation directions become available within minutes of eruption onset from low-coronal EUV observations, and the LASCO speed is typically measured within 1–2 hours thereafter, whereas in-situ speeds are only available upon CME arrival at Earth, at which point forecasting is no longer relevant. 

Correlations based on initial (near-Sun) CME speeds are generally expected to be lower than those using in-situ speeds, because CMEs undergo significant acceleration or deceleration during interplanetary transit \citep{gopalswamy2001predicting}. Indeed, previous studies using in-situ ICME speeds have reported correlations with Dst in the range \(r \approx 0.6\)–\(0.7\) \citep{srivastava2004solar, yurchyshyn2004correlation}, whereas \citet{yurchyshyn2004correlation} obtained a moderate correlation of only \(r \approx 0.53\) between CME initial speed and Dst for halo CMEs. The fact that our combined model achieves \(r = 0.7\) using initial speeds is therefore notable and suggests that the inclusion of the angular miss distance \(E\), which captures whether the CME nose or flank encounters Earth, compensates for the additional scatter introduced by using the near-Sun speed. In direct comparison, our model achieves \(r = 0.58\) using initial speed alone and \(r = 0.7\) when the DIRECD-derived angular miss distance is included. The improvement demonstrates that directional information partially compensates for the absence of in-situ data, making our correlation coefficients comparable to those achieved by \citet{srivastava2004solar} using in-situ speeds, despite relying exclusively on remote-sensing parameters available within hours of eruption onset.

More broadly, \citet{richardson2011geoeffectiveness} demonstrated that the southward component of the interplanetary magnetic field (\(B_z\)) remains the dominant driver of storm intensity, with CME speed playing a secondary modulating role, and that correlations based solely on CME speed rarely exceed \(r \approx 0.6\)–\(0.7\) unless magnetic field information is incorporated. Using in-situ parameters, \citet{kim2010empirical} developed an empirical Dst prediction model based on solar wind speed and \(B_z\), achieving prediction efficiencies of approximately 0.6–0.7 for moderate-to-intense storms, later refined for two-step forecasting by \citet{kim2014two}. 

Beyond single-event dynamics, \citet{rodriguez2020clustering, vennerstrom2016extreme} showed that successive CME impacts systematically enhance storm intensity beyond what single-event models predict. A comparable forecasting framework for a different class of geomagnetic storms was developed by \citet{Nitti2023}, who showed that CIR/HSS-driven storms can be predicted from coronal hole areas and magnetic polarity, achieving correlations of \(r = 0.63\)–\(0.73\) for the Dst index with a lead time of several days. That approach targets the recurrent, slowly evolving component of geomagnetic activity, whereas the present work addresses the impulsive, CME-driven component. Crucially, coronal dimmings are available within minutes of eruption onset, well before the CME enters the coronagraph field of view. This opens a complementary pathway for early geomagnetic storm forecasting from solar EUV observations, extending warning times for CME-driven events in a manner analogous to what coronal hole observations provide for CIR/HSS-driven storms.

A key additional outcome is the demonstrated potential for near real-time operation. For events processed using SDO ``quick-look'' data, the DIRECD pipeline successfully reproduced the full workflow, indicating that dimming-based CME direction estimation can be executed using promptly available EUV observations. This capability is relevant for operational space weather settings, where early estimates of CME direction and angular extent are needed as inputs to heliospheric propagation and ensemble forecasting tools, and where rapid updates can improve warning lead times.

Future extension of the catalog and continued development of the DIRECD software will focus on expanding its compatibility with near real-time data from additional EUV imagers (e.g., STEREO and future missions) to enhance its potential for operational space weather forecasting. In addition, a promising direction for future work is the extension of the DIRECD concept to off-limb EUV observations, which would allow dimming-based constraints on CME geometry to be applied to eruptions originating near or beyond the solar limb. Such developments would further strengthen the linkage between low-coronal signatures and coronagraph observations, enabling more continuous tracking of CME direction and expansion from the low corona into the outer corona. The public availability of both the catalog and software aims to encourage community adoption, validation, and collaborative improvement of this tool for CME direction analysis. As the catalog grows, it can also support statistically robust investigations of how low-coronal direction and geometry relate to CME evolution at larger LASCO/STEREO heights and to heliospheric consequences, strengthening the end-to-end connection from eruption onset to space weather impacts.

\begin{acknowledgments}
SDO data is courtesy of NASA/SDO and the AIA, and HMI science teams. The Large Angle Spectroscopic Coronagraph (LASCO) is a three coronagraph package which has been jointly developed for the Solar and Heliospheric Observatory (SOHO) mission by the Naval Research Laboratory (USA), the Laboratoire d'Astronomie Spatiale (France), the Max-Planck-Institut für Aeronomie (Germany), and the University of Birmingham (UK). We thank the referee for their valuable comments.
\end{acknowledgments}

\begin{contribution}
S.J. and T.P. developed the method and led the writing of the paper. A.V., K.D., and A.R. contributed to the conceptualization of this study, data analysis, and writing. S.J. developed the DIRECD software. All authors discussed the results and provided feedback on the manuscript.
\end{contribution}

\bibliography{bibliography}{}
\bibliographystyle{aasjournalv7}

\appendix

\section{Robustness check using the full 64-event catalog}
\label{appendixA}

\begin{figure}[h]
\centering
\includegraphics[width=\textwidth]{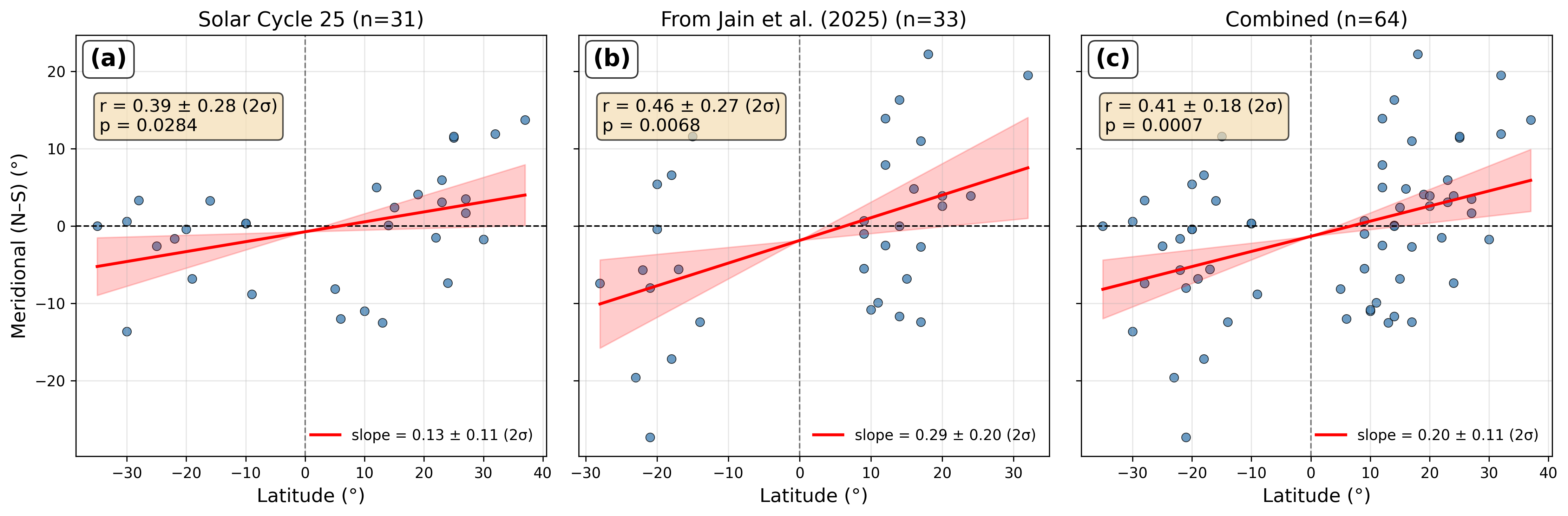}
\caption{Meridional inclination as a function of heliographic latitude for the Solar Cycle~25 sample (n=31), the legacy sample (n=33), and the full combined catalog (n=64).}
\label{fig:robustness}
\end{figure}

To test the robustness of the meridional inclination--latitude relation reported in Figure \ref{fig:summary_table} in Section~\ref{catalog}, we repeat the analysis after incorporating the 33 legacy events adopted from \citep{jain2025validating}, increasing the sample to 64 events. We apply the same linear regression and bootstrap correlation procedure described in Section~\ref{catalog} to three samples as shown in figure \ref{fig:robustness}: the 31 newly analyzed Solar Cycle~25 events alone (panel (a), same as panel (c) of Figure \ref{fig:summary_table}) , the 33 events from \citep{jain2025validating} (panel (b)), and the full combined catalog (panel (c)).

The meridional (N--S) deflection shows a significant positive trend with source latitude in all three samples: $r = 0.39$ ($p = 0.03$) for the Solar Cycle~25 sample alone , $r = 0.46$ ($p = 0.007$) for the legacy subsample alone, and $r = 0.41$ ($p < 0.001$) for the combined 64-event catalog. The trend is seen independently in both subsamples and it strengthens once the samples are combined, with the bootstrap uncertainty on $r$ shrinking from $\pm0.28$ (Solar Cycle~25 alone) to $\pm0.19$ (combined). 

\end{document}